\documentclass[lettersize,journal]{IEEEtran}

\usepackage{amsmath,amssymb,amsfonts}
\usepackage{graphicx}
\usepackage{booktabs}
\usepackage{multirow}
\usepackage{tabularx}
\usepackage{array}
\usepackage{url}
\usepackage{textcomp}
\usepackage{cuted}
\usepackage{capt-of}
\usepackage{pifont}
\usepackage{float}
\usepackage{dblfloatfix}
\usepackage[hidelinks]{hyperref}
\usepackage{orcidlink}

\usepackage{algorithm}
\usepackage{algpseudocode}
\usepackage{placeins}
\usepackage{microtype}
\usepackage[subtle]{savetrees}
\usepackage[dvipsnames]{xcolor}
\usepackage{enumitem}
\usepackage{cite}
\usepackage[caption=false,font=normalsize,labelfont=sf,textfont=sf]{subfig}
\usepackage{verbatim}
\usepackage[hidelinks]{hyperref}

\newcommand{\cmark}{\ding{51}}

\begin{document}

\title{Beyond Centralized Policy Decision Points: \\ Decentralized Sticky Policy Authorization \\ through Evidence Quorums}

\author{
Aleena Nazir\,\orcidlink{0009-0007-9499-4999},
Mubashir Husain Rehmani\,\orcidlink{0000-0002-3565-7390},
Bernard Butler\,\orcidlink{0000-0003-0610-0128},
and Donna O'Shea\,\orcidlink{0000-0002-2437-3106}
\thanks{Aleena Nazir is with the Bernal Institute, University of Limerick,
Ireland (e-mail: nazir.aleena@ul.ie).}
\thanks{Mubashir Husain Rehmani is with Munster Technological University,
Ireland (e-mail: mshrehmani@gmail.com).}
\thanks{Bernard Butler is with South East Technological University,
Ireland (e-mail: bbutler@ieee.org).}
\thanks{Donna O'Shea is with the Bernal Institute, University of Limerick, Ireland (e-mail: donna.c.oshea@ul.ie).}
\thanks{Corresponding authors: Aleena Nazir and Donna O'Shea.}
}



\maketitle

\begin{abstract}
Sticky policies remain attached to protected data so that access restrictions can persist across systems, yet request-time authorization often still depends on a centralized Trusted Authority or Policy Decision Point (PDP). This paper presents SPEAR-Q, a decentralized framework in which independent Policy Authority Nodes (PANs) evaluate the active sticky policy using local context, consent, and risk information and generate signed authorization evidence. A strict-majority Evidence quorum determines the global Permit or Deny decision, while a committed policy becomes active only after the required PAN majority applies it. SPEAR-Q was deployed across physically separated hosts in the Airbus CyberSecurity Simulation Platform and evaluated using a deterministic healthcare workload derived from dataset MIMIC-IV. Across 9,000 performance requests, all requests completed without execution failure or timeout. Latency increased and throughput saturated as cluster size and concurrency grew, with PAN evidence waiting and PAN-local processing dominating under load and credential validation forming the main PAN-side cost. Policy-activation experiments confirmed majority-based activation. Quorum experiments showed that correctly signed false-Permit evidence from fewer compromised PANs than the required quorum could not produce a global Permit or resource release, while authorization remained fail-closed when reachable PANs fell below the required quorum. Compared with prior approaches that decentralize policy management, credential authority, or threshold-based release separately, SPEAR-Q integrates persistent sticky policy enforcement, independent request-time evidence, majority-based policy activation, and explicit quorum-bounded authorization within one decentralized framework. Under the evaluated conditions, SPEAR-Q therefore supports decentralized sticky-policy authorization without relying on a centralized decision authority. 
\end{abstract}

\begin{IEEEkeywords}
Sticky policies, decentralized authorization, evidence quorum, access control, zero trust, fault tolerance.
\end{IEEEkeywords}


\section{Introduction}
\label{sec:introduction}

\IEEEPARstart{S}{ticky} policies are machine-readable access control policies that remain attached to the data they govern, allowing restrictions on access and use to persist as data moves beyond its original system or administrative domain. In simple terms, an access control policy defines who can access a protected resource, what actions they are allowed to perform, and under what conditions that access is permitted. While this ensures that policy information travels with the data, sticky policies do not themselves determine whether an access request should be permitted or denied. Policy evaluation and authorization decisions are typically performed by a Trusted Authority or Policy Decision Point (PDP) responsible for interpreting the attached policy and generating an authorization outcome. As a result, sticky policy architectures remain dependent on a centralized decision making authority even though the policy accompanies the data. This dependency introduces a potential single point of failure and trust, whereby the unavailability or compromise of the authority can disrupt authorization services or affect system security. This challenge is particularly relevant in distributed data-sharing environments where sensitive information moves across organizational and administrative boundaries.

Although this challenge applies across multiple domains, healthcare is used in this work as a motivating use case and evaluation setting in this paper. Modern healthcare ecosystems rely on interconnected IoT devices, electronic health records, and cloud-based analytics, with patient data increasingly moving across multiple systems and administrative domains. In such environments, request-time authorization determines whether a specific user or system is permitted to perform an action on protected data at the time the request is made, using the governing policy and the authorization information available at that moment. However, a valid authorization decision at one point of access does not necessarily control how the data is subsequently used, shared, or retained. Governance obligations such as purpose limitation, retention control, and auditability, therefore, need to extend beyond the point of access. This challenge is illustrated by a 2025 U.S. Department of Health and Human Services investigation involving BayCare Health System, where credentials associated with a healthcare practice with legitimate access for continuity of care were used to access a patient's electronic medical record. The patient was later contacted by an individual who had photographs of her printed records and a video of her electronic medical record \cite{hhs2025baycare}. The incident demonstrates that legitimate access does not necessarily ensure that subsequent data use remains consistent with governing policies.

To maintain policy restrictions as data moves across systems, prior work has incorporated sticky policies into distributed access-control architectures. Sicari et al. show that sticky policies can move with data and support distributed enforcement, but the Policy Decision Point remains within an external Trusted Authority, making authorization dependent on an always-available central entity \cite{sicari2021attribute}. The authors identify this Trusted Authority as a potential bottleneck and single point of failure. Subsequent work by Rizzardi et al. removes the Trusted Authority by using a permissioned blockchain to distribute policy management and protect policy updates \cite{rizzardi2022securing}. However, its consensus mechanism is primarily used for policy creation, update, and revocation rather than for combining independently generated request-time authorization evidence. Thus, despite progress toward distributed policy enforcement and policy management, decentralized request-time authorization remains a distinct challenge. Section \ref{related_work} provides a broader comparison of the state-of-the-art approaches relevant to sticky policy enforcement and decentralized authorization.

To address this gap, we present Sticky Policy Enforcement with Authority Nodes using Quorum-based Authorization (SPEAR-Q), a decentralized framework that provides policy decision and enforcement without relying on a centralized Trusted Authority. Requests are forwarded through an active-active gateway layer, in which multiple gateways remain operational simultaneously and can independently accept and forward authorization requests, allowing traffic to continue through another gateway if one becomes unavailable. The gateways do not make authorization decisions; they only provide redundant request paths to the PAN cluster. Requests are then forwarded to multiple Policy Authority Nodes (PANs), which independently interpret and evaluate the sticky policy using locally available context, consent, and user-risk information and generate signed authorization evidence. The Resource Verifier validates this evidence and derives the global Permit or Deny decision through a strict-majority Evidence quorum, ensuring that no individual PAN determines access alone. A Permit allows the protected data to be released, while a Deny prevents its release. Because the sticky policy remains associated with the data, any later access or sharing request is evaluated again against the governing policy before the data is released further. By re-evaluating the sticky policy for subsequent access or sharing requests and retaining the resulting authorization record, SPEAR-Q extends policy enforcement beyond the initial access decision, allowing governing restrictions and auditability to remain effective as data moves across participating systems.

The security and resilience of this authorization model are also examined under an explicit threat model covering manipulated policy and evidence, replay, PAN failure and compromise, and network disruption. This evaluation assesses whether authorization remains reliable under failure and adversarial conditions while measuring system performance and concurrency.

The main contributions of this work are summarized as follows:

\begin{itemize}

\item Present SPEAR-Q, a decentralized framework that provides both sticky policy decision and enforcement through an active-active gateway layer and independent Policy Authority Nodes (PANs), removing reliance on a centralized Trusted Authority for request-time authorization.

\item Develop an Evidence quorum authorization model in which independently
generated PAN evidence is combined through a strict-majority decision,
defining an explicit compromise bound for correctly signed false-Permit
evidence and fail-closed behaviour when sufficient valid evidence is
unavailable.

\item Implement SPEAR-Q on physically separated Airbus cyber-range hosts and evaluate performance, concurrency, authorization cost, policy activation, Evidence-quorum behavior, and post-authorization processing using MIMIC-IV-derived healthcare workloads.

\end{itemize}

The remainder of this paper is organized as follows.
Section \ref{related_work} reviews related work.
Section \ref{sec:methodology} presents the system model, threat model, and authorization mechanisms. Section \ref{sec:experimentation} presents the evaluation, results, and discussion. Section~\ref{sec:conclusion} concludes the paper.


\section{Related Work}
\label{related_work}

Research related to SPEAR-Q spans four complementary directions:
sticky policy and usage-control enforcement, blockchain-based
decentralization, decentralized and multi-authority authorization, and
quorum-based fault tolerance. These approaches distribute different
parts of the access-control path. Some preserve policy state with the
data, others decentralize policy or credential management, and others
require agreement among multiple participants. We distinguish these approaches according to where authorization authority resides, whether multiple parties participate in the same request-time decision, and whether explicit failure or compromise bounds are considered.

\subsection{Sticky Policies and Persistent Usage Control}

Sticky policies associate machine-readable usage constraints with data
so that the applicable policy can persist when the data crosses system
or administrative boundaries. Sicari et al.
\cite{sicari2017security} apply sticky policies to Networked Smart
Objects, while Miorandi et al. \cite{miorandi2019sticky} characterize
the broader design space and lifecycle of sticky policy enforcement.
More recent work extends this principle to GDPR-aware cloud data
sharing \cite{cambronero2024towards} and combines sticky policies with
Attribute-Based Encryption (ABE) to protect data disclosure
\cite{sicari2021attribute}. Usage-control and data-space approaches
similarly express persistent constraints governing how data can be
accessed, processed, and shared
\cite{akaichi2022usage,dam2023policy,maamar2023make,da2025data}.

These systems establish an important data-centric property: policy
state can remain associated with the protected object rather than being
defined only at an application boundary. This property, however, is
orthogonal to the placement of the runtime authorization authority.
A policy can travel with data while its interpretation, validation, or permission issuance still depends on a centralized Trusted Authority or PDP.

\subsection{Blockchain-Based Decentralized Access Control}

Blockchain is widely used to reduce reliance on centralized
access-control infrastructure by distributing policy state, transaction
validation, and audit records. Hasan et al. \cite{hasan2023smart}
encode IoT access-control logic in smart contracts, while Khalid
et al. \cite{khalid2023towards} combine software-defined networking
with smart-contract-based access control. In healthcare, Pu
et al. \cite{pu2024medical} employ smart contracts for medical-data
access control, and Karankar and Seth \cite{karankar2025iot} combine
blockchain with messaging mechanisms for distributed IoT access
management. Such systems reduce dependence on a single policy-management
service and provide tamper-evident coordination of access-control state.

Rizzardi et al. \cite{rizzardi2022securing} are the closest work to the problem considered in this paper, using a permissioned blockchain to remove the centralized Trusted Authority from sticky policy creation, update, and revocation. However, their decentralization is applied to maintaining and protecting the sticky policy rather than to the request-time authorization decision itself. When a resource request is made, the authorization outcome is not derived from independently generated decisions or signed authorization evidence from multiple authorities. Thus, while policy maintenance is decentralized, the problem of decentralized request-time authorization for each policy-bound request remains unresolved.

A recent NDN-based decentralized authorization framework for medical data protection by Alshahrani et al. \cite{alshahrani2026decentralized} moves closer to request-level decentralized authorization. Their NDN-based framework combines
smart-contract-based access control with proxy re-encryption and uses
a Byzantine-fault-tolerant validator model. Validators independently
evaluate the applicable smart-contract policy, produce signed binary
votes, and a $2f+1$ quorum determines the access outcome. The approach therefore demonstrates that request-time authorization can be derived through multi-party agreement rather than a single PDP.

\begin{table*}[!t]
\caption{Comparison of representative work on persistent policy
enforcement, decentralized access control, and quorum-based
authorization. SP: implemented sticky policy enforcement;
DA: decentralized request-time authorization authority;
Req.-Q: explicit multi-party agreement or threshold used for the same
authorization or protected operation; FB: explicit distributed fault or
compromise bound governing the authorization decision; FE: empirical
failure of the authorization path.
\ding{51}: explicitly supported;
$\triangle$: partially or indirectly supported;
\ding{55}: not supported;
--: not applicable to the scope of the cited work.}
\label{tab:related_comparison}

\centering
\scriptsize
\setlength{\tabcolsep}{3pt}
\renewcommand{\arraystretch}{1.08}

\begin{tabularx}{\textwidth}{|p{2.7cm}|X|c|c|c|c|c|}
\hline

\textbf{Work} &
\textbf{Primary Focus / Mechanism} &
\textbf{SP} &
\textbf{DA} &
\textbf{Req.-Q} &
\textbf{FB} &
\textbf{FE} \\
\hline

\multicolumn{7}{|l|}{\textit{Sticky policies, usage control, and policy-based enforcement}} \\
\hline

Sicari et al. \cite{sicari2017security} (2017) &
Sticky policy enforcement for Networked Smart Objects with external authorization authority &
\ding{51} & \ding{55} & \ding{55} & \ding{55} & \ding{55} \\
\hline

Miorandi et al. \cite{miorandi2019sticky} (2019) &
Survey of sticky policy models, architectures, and lifecycle enforcement &
-- & -- & -- & -- & -- \\
\hline

Sicari et al. \cite{sicari2021attribute} (2021) &
ABE combined with sticky policies for cryptographically protected data access &
\ding{51} & \ding{55} & \ding{55} & \ding{55} & \ding{55} \\
\hline

Caserio et al. \cite{caserio2022formal} (2022) &
Formal validation of XACML 3.0 access-control policies &
-- & -- & -- & -- & -- \\
\hline

Akaichi and Kirrane \cite{akaichi2022usage} (2022) &
Survey of usage-control specification, enforcement, and robustness &
-- & -- & -- & -- & -- \\
\hline

Dam et al. \cite{dam2023policy} (2023) &
Usage-control policy patterns for data spaces &
-- & -- & -- & -- & -- \\
\hline

Maamar et al. \cite{maamar2023make} (2023) &
ODRL-based privacy and usage-policy representation for IoT &
-- & -- & -- & -- & -- \\
\hline

Cambronero et al. \cite{cambronero2024towards} (2024) &
GDPR-oriented cloud architecture using policies attached to data &
\ding{51} & \ding{55} & \ding{55} & \ding{55} & \ding{55} \\
\hline

Defersha \cite{defersha2025privacybridge} (2025) &
Sticky policy-based control for GenAI data governance &
\ding{51} & \ding{55} & \ding{55} & \ding{55} & \ding{55} \\
\hline

da Gama Cordeiro and de Oliveira \cite{da2025data} (2025) &
Data access and usage-control mechanisms for data spaces &
-- & -- & -- & -- & -- \\
\hline

\multicolumn{7}{|l|}{\textit{IoT and application-specific access control}} \\
\hline

Ragothaman et al. \cite{ragothaman2023access} (2023) &
Survey of dynamic IoT access-control policies and research directions &
-- & -- & -- & -- & -- \\
\hline

Trabelsi et al. \cite{trabelsi2023access} (2023) &
Survey of IoT access-control models and architectures &
-- & -- & -- & -- & -- \\
\hline

Shojaei et al. \cite{shojaei2024security} (2024) &
Systematic review of security and privacy in health information systems &
-- & -- & -- & -- & -- \\
\hline

Alaba et al. \cite{alaba2025iot} (2025) &
Review of IoT applications, security challenges, and healthcare deployments &
-- & -- & -- & -- & -- \\
\hline

Kalaria et al. \cite{kalaria2024adaptive} (2024) &
Adaptive context-aware XACML enforcement using fog computing &
\ding{55} & $\triangle$ & \ding{55} & \ding{55} & \ding{55} \\
\hline

Vijayaraghavan \cite{vijayaraghavan2025policy} (2025) &
Policy-as-Code approach for infrastructure security &
-- & -- & -- & -- & -- \\
\hline

\multicolumn{7}{|l|}{\textit{Blockchain-based decentralized access control}} \\
\hline

Rizzardi et al. \cite{rizzardi2022securing} (2022) &
Consensus-based sticky policy creation, update, and revocation;
no independent evidence quorum for each request-time authorization decision &
\ding{51} & $\triangle$ & \ding{55} & \ding{55} & \ding{55} \\
\hline

Hasan et al. \cite{hasan2023smart} (2023) &
Smart-contract-based IoT access-control framework &
\ding{55} & \ding{51} & $\triangle$ & \ding{55} & \ding{55} \\
\hline

Khalid et al. \cite{khalid2023towards} (2023) &
Blockchain smart contracts integrated with SDN-based IoT access control &
\ding{55} & $\triangle$ & \ding{55} & \ding{55} & \ding{55} \\
\hline

Pu et al. \cite{pu2024medical} (2024) &
Blockchain smart-contract and risk-based medical access control &
\ding{55} & \ding{51} & $\triangle$ & \ding{55} & \ding{55} \\
\hline

Li et al. \cite{li2024damfsd} (2024) &
Blockchain-based decentralized authorization and permission delegation for medical data &
\ding{55} & \ding{51} & \ding{55} & \ding{55} & \ding{55} \\
\hline

Karankar and Seth \cite{karankar2025iot} (2025) &
Blockchain, ABAC, smart contracts, and message queuing for decentralized IoT access control &
\ding{55} & \ding{51} & $\triangle$ & \ding{55} & \ding{55} \\
\hline

Alshahrani et al. \cite{alshahrani2026decentralized} (2026) &
SCBAC/PRE with independent BFT-validator authorization voting &
\ding{55} & \ding{51} & \ding{51} & \ding{51} & $\triangle$ \\
\hline

\multicolumn{7}{|l|}{\textit{Multi-authority, cryptographic, and threshold authorization}} \\
\hline

Shafagh et al. \cite{shafagh2020droplet} (2020) &
Decentralized authorization service for encrypted IoT data streams &
\ding{55} & \ding{51} & \ding{55} & \ding{55} & \ding{55} \\
\hline

Zichichi et al. \cite{zichichi2020personal} (2020) &
Distributed authorization using secret sharing and threshold data release &
\ding{55} & \ding{51} & \ding{51} & \ding{55} & \ding{55} \\
\hline

Wu et al. \cite{wu2025decentralised} (2025) &
Decentralized multi-authority ABE for IoT access control &
\ding{55} & $\triangle$ & \ding{55} & \ding{55} & \ding{55} \\
\hline

Li et al. \cite{li2025decentralized} (2025) &
Decentralized healthcare delegation requiring a threshold of collaborating doctors &
\ding{55} & \ding{51} & \ding{51} & \ding{55} & \ding{55} \\
\hline

Wang et al. \cite{wang2026decentralized} (2026) &
Decentralized multi-authority ABE for eHealthcare &
\ding{55} & $\triangle$ & \ding{55} & \ding{55} & \ding{55} \\
\hline

\multicolumn{7}{|l|}{\textit{Quorum, replication, and distributed fault-tolerance foundations}} \\
\hline

Mondal et al. \cite{mondal2022applying} (2022) &
Secure heterogeneous quorum replication with bounded faults &
-- & -- & \ding{51} & \ding{51} & -- \\
\hline

Zhang and Tseng \cite{zhang2024fault} (2024) &
Fault-tolerant consensus in dynamic distributed networks &
-- & -- & -- & -- & -- \\
\hline

Tennage et al. \cite{tennage2025racs} (2025) &
Robust randomized consensus for cloud systems &
-- & -- & -- & -- & -- \\
\hline

Naser-Pastoriza et al. \cite{naser2025tight} (2025) &
Theoretical bounds for quorum systems &
-- & -- & -- & -- & -- \\
\hline

\multicolumn{7}{|l|}{\textit{Proposed framework}} \\
\hline

\textbf{SPEAR-Q} &
\textbf{Independent sticky policy evaluation with Evidence quorum
authorization and separate replicated coordination} &
\ding{51} &
\ding{51} &
\ding{51} &
\ding{51} &
\ding{51} \\
\hline

\end{tabularx}
\end{table*}

\subsection{Decentralized and Multi-Authority Authorization}

Decentralized authorization can also be achieved without blockchain
consensus. Droplet \cite{shafagh2020droplet} provides a decentralized
authorization service for encrypted data streams, coupling
fine-grained access policies with cryptographic protection so that data
owners need not rely on the storage provider for access control.
DAMFSD \cite{li2024damfsd} similarly addresses the single-authority
problem in medical-data sharing through decentralized and transitive
permission delegation. These systems demonstrate that authorization
and delegation services can be distributed independently of the
storage infrastructure.

Multi-authority cryptographic schemes decentralize another component
of the authorization process. Decentralized and multi-authority ABE
systems distribute attribute or key issuance across several
authorities, reducing key escrow and reliance on a single credential
issuer \cite{wu2025decentralised,wang2026decentralized}. Access is
ultimately obtained by satisfying ciphertext-embedded cryptographic
conditions using credentials issued by the participating authorities.
Thus, multiple authorities contribute to the establishment of
authorization capabilities, but they need not independently evaluate
the same request at the time a resource is released.

\subsection{Quorum-Based and Fault-Tolerant Authorization}

Threshold and quorum mechanisms provide another route to eliminating
unilateral decision authority. Zichichi et
al. \cite{zichichi2020personal} combine distributed authorization with
secret sharing and threshold proxy re-encryption, requiring
participation from multiple entities before protected personal data can
be disclosed. Their work demonstrates threshold-controlled data
release, but does not address persistent sticky policy evaluation or
the behavior of the authorization service under explicit node-failure
scenarios.

At a more general systems level, FLAQR
\cite{mondal2022applying} formalizes secure quorum replication for
distributed applications. It requires a quorum of replicated hosts to
agree on operations and provides confidentiality, integrity,
availability, and liveness properties under bounded failures.
Complementary work studies robust consensus, quorum-system bounds, and
fault-tolerant agreement in distributed environments
\cite{tennage2025racs} \cite{naser2025tight,zhang2024fault}.
These works provide foundations for reasoning about agreement and replication, but do not define the authorization semantics of a sticky policy-protected resource request.

\subsection{Research Gap and Positioning of SPEAR-Q}
Taken together, existing approaches distribute different components of the access-control process, but do not fully combine persistent sticky policy enforcement with decentralized request-time authorization. Sticky policy approaches preserve policy requirements with the data, blockchain-based systems primarily decentralize policy management or contract execution, multi-authority schemes distribute credential or key authority, and threshold mechanisms require multiple parties for selected authorization or release operations. However, these properties are generally addressed separately rather than as a request-level authorization model based on independently generated authorization evidence.

SPEAR-Q addresses this gap by combining sticky policy enforcement with decentralized request-time authorization through independently generated PAN evidence and an Evidence quorum. This defines an explicit compromise bound under which correctly signed
false-Permit evidence from a sub-quorum coalition cannot independently
authorize a request, while supporting fail-closed authorization under
insufficient evidence. Because policy updates may reach distributed PANs at different times, SPEAR-Q also uses versioned policy activation and PAN eligibility checks to ensure that only PANs aligned with the active policy state contribute authorization evidence, as described in Section \ref{sec:policy_sync}. Raft is used only for policy-update commitment. Table \ref{tab:related_comparison} summarizes this positioning using the technical properties relevant to the identified gap.


\section{Methodology}
\label{sec:methodology}

This section presents the technical design and operation of SPEAR-Q. It defines the system architecture, authorization flow, threat model and security properties, followed by sticky policy update and activation, independent PAN evidence generation, Evidence quorum authorization, and authorization commitment with PAN-local risk evolution.


\subsection{System Model}
\label{sec:system_model01}

The system model defines the entities participating in SPEAR-Q. The System Architecture defines the role of each entity, while the Authorization Flow summarizes their interaction during policy management, request-time authorization, and post-authorization coordination before the detailed mechanisms are presented in the following subsections.


\subsubsection{System Architecture}
\label{sec:system_model}

\begin{figure*}[t]
\centering
\includegraphics[width=0.98\textwidth]{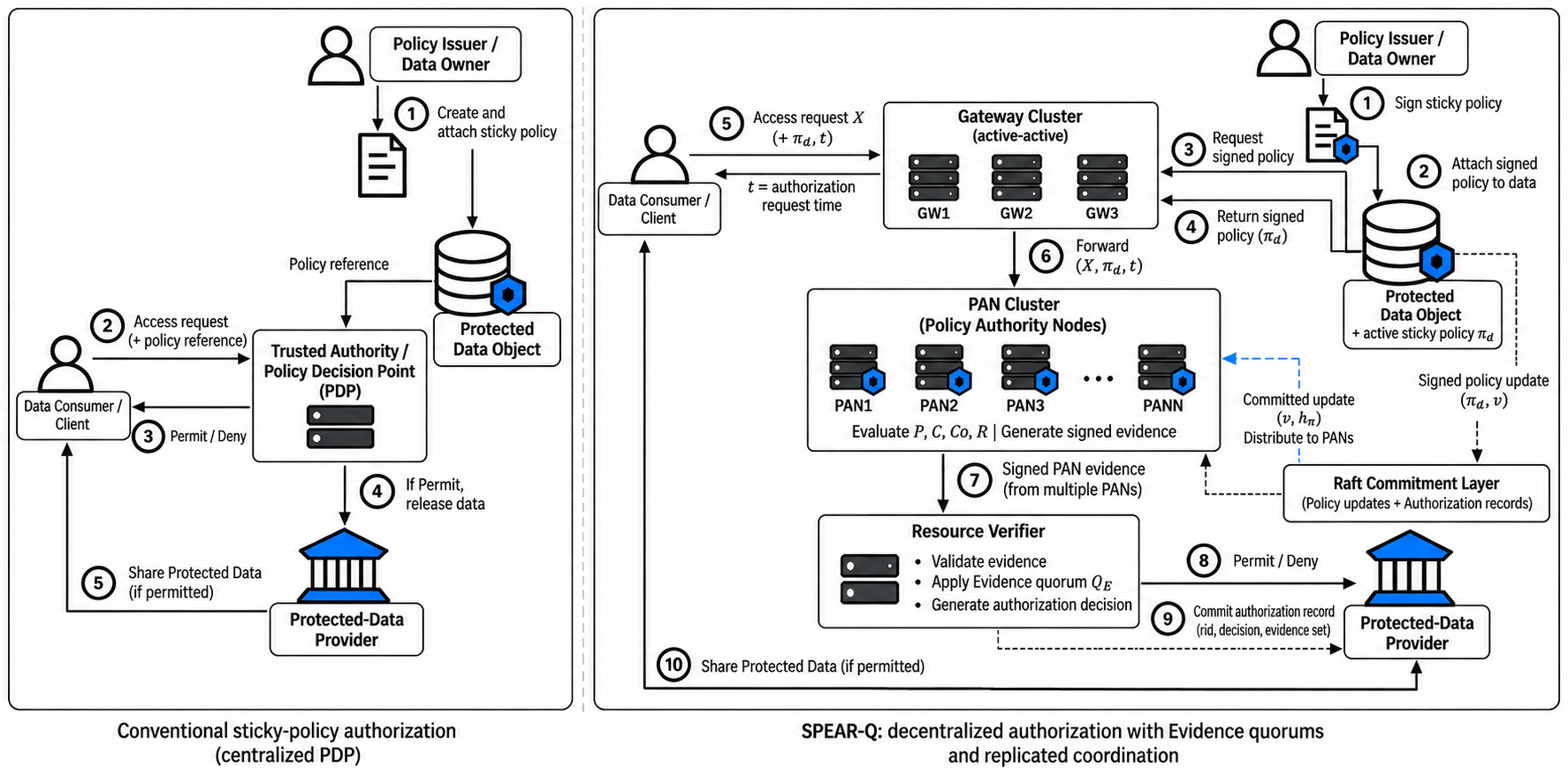}
\caption{Comparison of sticky policy authorization models.
Left: conventional authorization through a centralized Trusted Authority or
Policy Decision Point (PDP). Right: SPEAR-Q, where PAN Cluster independently evaluate the sticky policy and the Resource Verifier
derives the final decision from signed PAN evidence through an Evidence
quorum.}
\label{fig:model_comparison}
\end{figure*}

SPEAR-Q comprises a set of cooperating entities that collectively support policy management, request-time authorization, evidence validation, authorization commitment, and protected-resource release. The architecture consists of a Data Consumer, Gateway Cluster, PAN Cluster, Resource Verifier, Protected-Data Provider, Policy Issuer/Data Owner, and Raft commitment layer. Each protected data object \(d\) remains associated with a versioned sticky policy \(\pi_d\). The Policy Issuer/Data Owner creates and signs these policies, while the Raft commitment layer records validated policy updates and resulting authorization records. Request-time authorization is derived from independently generated evidence across the distributed PAN Cluster. Figure~\ref{fig:model_comparison} presents the overall architecture and illustrates the roles and interactions of its principal components. The role of each component is described below.

\textbf{Data Consumer:}
The Data Consumer initiates an authenticated request to perform action $a$ on
protected data $d$. The request is represented as
$X=(rid,role,d,a,t,\ell)$, where $rid$ uniquely identifies the request,
$role$ denotes the user role, $a$ represent action, $t$ and $\ell$ represent the request
time and location. The authenticated session additionally associates the
request with user identity $u$, which is used to index user-specific
consent and risk state during PAN evaluation.

\textbf{Gateway Cluster:}
The Gateway Cluster provides the entry point for authorization requests and uses concurrently active gateways to provide redundant request paths. If one gateway becomes unavailable, requests can continue through another reachable gateway. It does not evaluate authorization conditions or contribute authorization evidence.

\textbf{Policy Authority Nodes:} The PAN Cluster contains \(N\geq3\) Policy Authority Nodes. The evaluated
configurations use \(N\in\{3,5,7\}\). Each PAN maintains a local information base containing its applied policy state, validated context information, validated consent information for the requested data and action, and PAN-local user-risk history. PANs use the same active policy state, while their local context, consent, and risk information may differ. Each PAN therefore evaluates the request independently using its locally available authorization information.

A PAN contributes authorization evidence only when its locally applied policy version and digest match the active policy associated with the request. Eligible PANs return independently generated signed evidence for subsequent quorum processing. Policy update and activation are defined in section~\ref{sec:policy_sync}, while PAN eligibility, local evaluation, and evidence generation are formalized in algorithm \ref{alg:evidence_generation}.

\textbf{Resource Verifier:}
The Resource Verifier collects signed PAN evidence, validates its admissibility, and applies the Evidence quorum rule to derive the global authorization decision. It does not act as an independent authorization authority: a Permit can be derived only when a strict majority of distinct admissible PAN records support Permit, and the verifier cannot substitute for missing PAN evidence or contribute its own authorization vote.  Thus, authorization authority remains distributed across the PAN Cluster even though the Resource Verifier coordinates evidence validation and derivation of the final decision.

\textbf{Protected-Data Provider:}
The Protected-Data Provider holds the protected resource and releases it only after validating the authorization certificate associated with a committed global Permit. A Deny, failed commitment, or absence of a valid authorization certificate results in no resource release.

\textbf{Policy Issuer:}
The Policy Issuer, which may also be the Data Owner, creates and signs the versioned sticky policy associated with each protected data object. The issuer signature identifies the authorized origin of a policy update, while the associated version and digest identify the policy state.

\textbf{Raft Commitment Layer:}
The Raft commitment layer maintains ordered commitment of validated policy updates and resulting authorization records. It does not evaluate authorization conditions, generate PAN evidence, or alter the global authorization decision.

\subsubsection{Authorization Flow}
\begin{figure}[t]
\centering
\includegraphics[width=\linewidth]{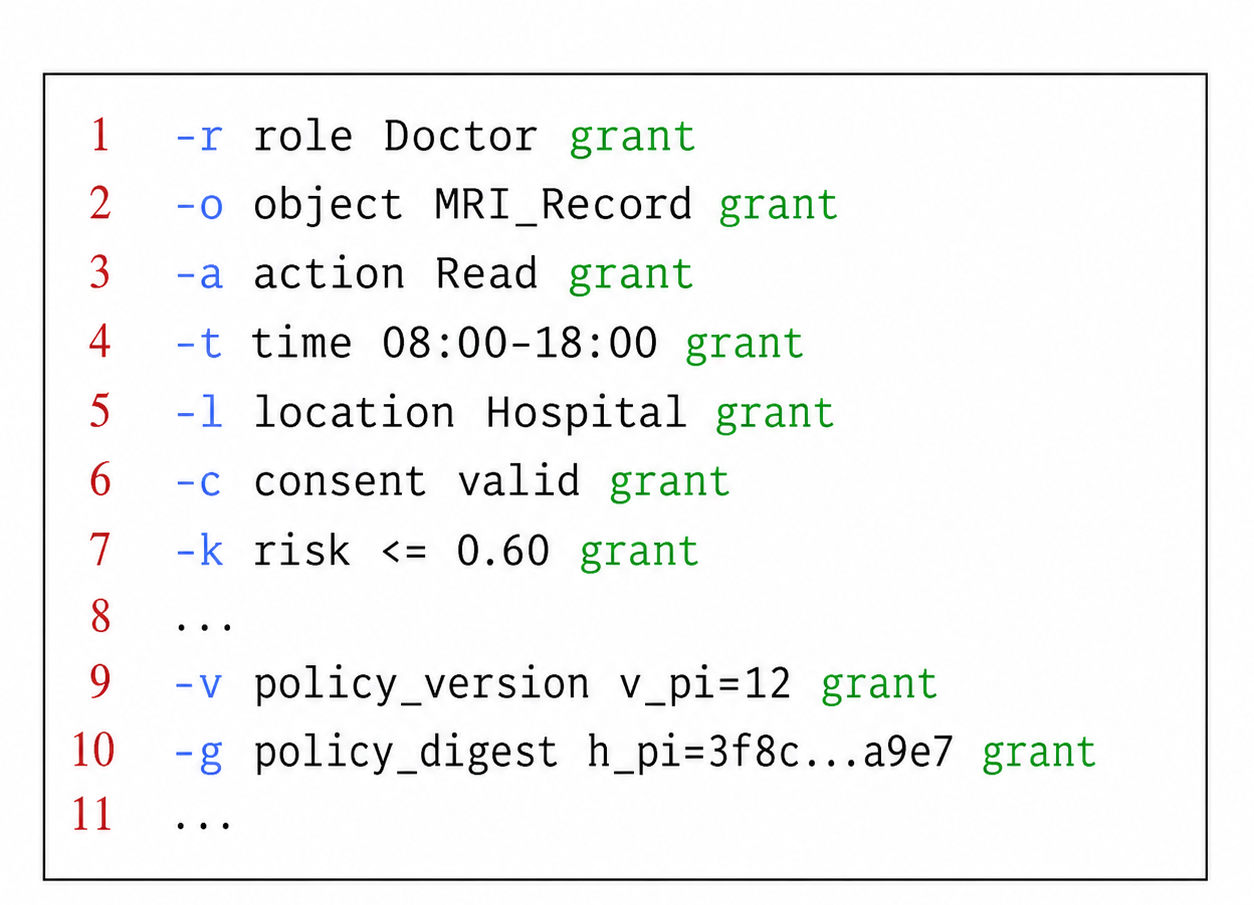}
\caption{Example sticky policy representation used for a healthcare data in SPEAR-Q.}
\label{fig:sticky}
\end{figure}

The authorization flow comprises policy management, request-time authorization, and post-authorization coordination. A signed policy update from the Policy Issuer/Data Owner is validated, committed through Raft, and applied across the PAN Cluster before it becomes the active policy state for subsequent requests. Until this activation completes, the preceding active policy remains in use.
Figure~\ref{fig:sticky} illustrates the sticky policy fields used during
authorization, including requester role (\(role\)), protected object (\(d\)), requested action (\(a\)), time (\(t\)), location (\(\ell\)), consent requirement, risk threshold (\(\rho\)), policy version (\(v_{\pi}\)), and policy digest (\(h_{\pi}\)). The policy version and digest identify the active policy state, allowing the PAN to verify that the received policy matches the version authorized for request-time evaluation.

Once an active policy is available, the Data Consumer submits the authorization request \(X\) through the Gateway Cluster. The Gateway Cluster retrieves the sticky policy \(\pi_d\) associated with the protected data and forwards the request-policy bundle \((X,\pi_d)\) to the PAN Cluster. Eligible PANs independently evaluate the request using their local authorization information and return signed evidence. The Resource Verifier validates the received evidence and applies the Evidence quorum \(Q_E\) to derive the global Permit or Deny decision.

\begin{figure}[t]
\centering
\includegraphics[width=\linewidth]{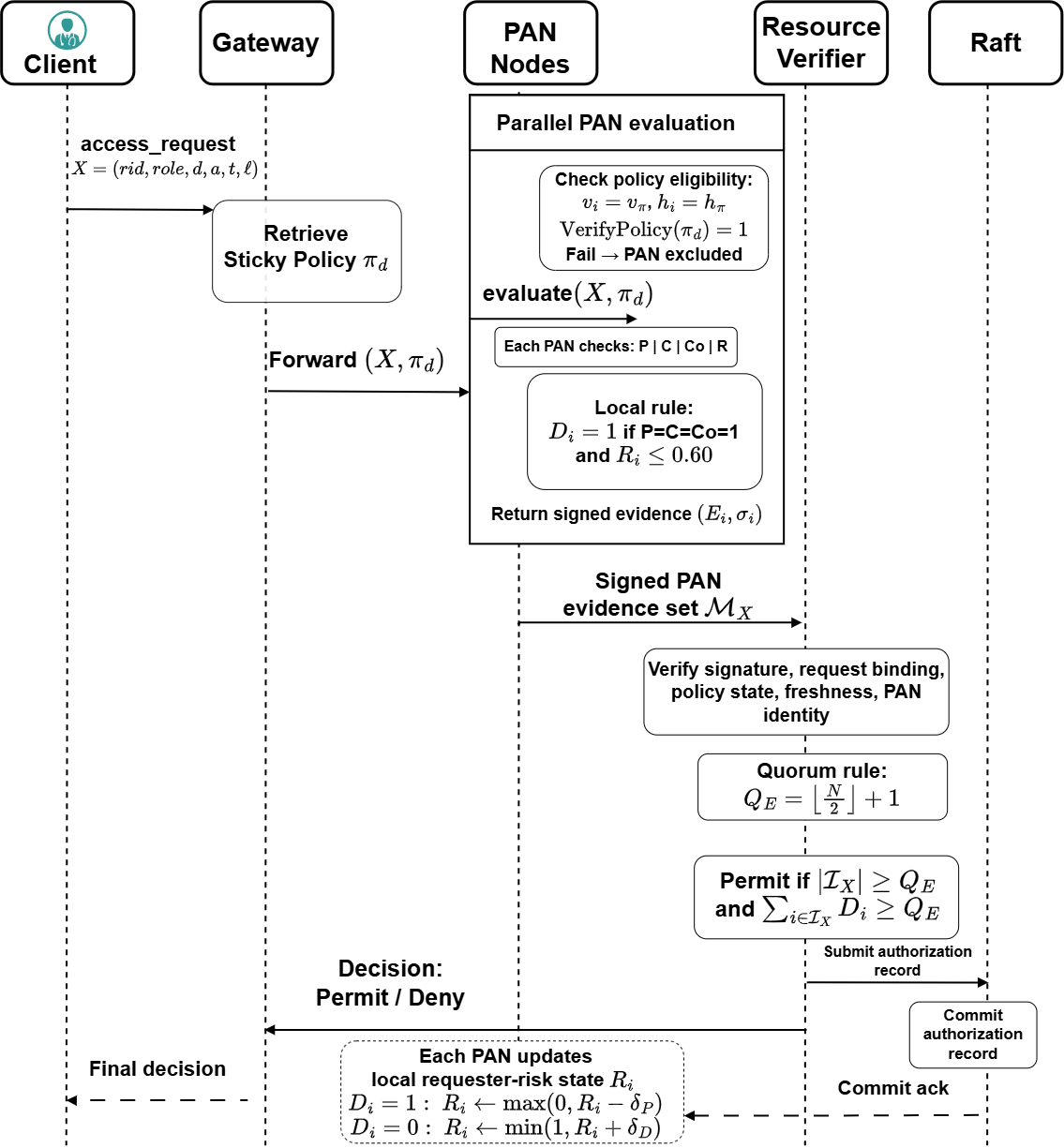}
\caption{SPEAR-Q authorization workflow}
\label{fig:protocol_flow}
\end{figure}

The resulting authorization record is then submitted to Raft for commitment. For a successfully committed global Permit, the Resource Verifier issues the authorization certificate required by the Protected-Data Provider, which validates the certificate before releasing the protected resource to the Data Consumer. A Deny or unsuccessful commitment results in no resource release. A committed authorization outcome may subsequently update the PAN-local risk state used in later request evaluations. Figure \ref{fig:protocol_flow} summarizes this request-time sequence. The following subsections formalize the distributed information state, policy update and activation procedure, PAN evidence generation, Evidence quorum authorization, and post-authorization commitment.

\subsection{Threat Model and Security Properties}
\label{sec:threat_model}

\textbf{Protected assets \& Adversary Objective: }
The protected assets are the protected data \(d\), the active sticky policy
reference, PAN authorization evidence, and the resulting authorization
decision. The adversary seeks to cause an unauthorized resource release by
manipulating the request, policy state, or authorization evidence available
during request-time authorization. The adversary may also disrupt communication to reduce the evidence available for quorum formation.

\textbf{Trust and attacker capabilities: }
The authorized Policy Issuer, Resource Verifier, and Protected-Data Provider are trusted. Gateways only forward the request-policy bundle and do not contribute authorization evidence. The network is adversarial, and messages may be delayed, dropped, duplicated, replayed, reordered, or modified; gateways or PANs may also become unavailable. PANs operate independently, and a subset may be compromised. A compromised PAN may withhold or submit malformed evidence, return correctly signed false evidence, or collude with others. Policy state may be stale, modified, mismatched, or substituted, while evidence may be modified, duplicated, expired, or replayed. Only policy-valid PANs and admissible evidence contribute to the quorum. Raft is considered only under crash and communication faults. It commits the authorization record but does not determine the global decision; commitment failure prevents resource release and PAN-local risk update.

\textbf{Compromised-PAN authorization bound:}
Let \(m\) denote the number of
compromised PANs. Because a Permit requires \(Q_E\) distinct admissible records carrying Permit support, a coalition with \(m<Q_E\) cannot independently satisfy the authorization rule. Likewise, loss or withholding of evidence cannot lower this requirement: if fewer than \(Q_E\) admissible records are available, authorization returns Deny. When \(m\geq Q_E\), the coalition can satisfy the Permit-support requirement; therefore, no authorization-safety guarantee is claimed at or above this bound. This bound limits the influence of compromised PANs but does not identify or detect
which PANs are malicious.



\subsection{Sticky Policy Updates and Activation}
\label{sec:policy_sync}
SPEAR-Q separates policy update from policy activation. An update introduces a candidate policy state, whereas activation determines when a successfully committed candidate replaces the active policy state used for subsequent authorization. Until a new version is successfully committed and activated, the previously active policy remains in use. Algorithm \ref{alg:policy_sync} formalizes the complete policy update and activation procedure, including validation, Raft commitment, PAN-local application, and activation.

Each protected data $d$ is associated with a versioned sticky policy $\pi_d$. Its metadata records the policy version, policy digest, update time, authorized issuer, and issuer signature. The policy digest is a SHA-256 cryptographic fingerprint of the encoded policy content and is used to verify the integrity of the received policy state.

When a signed policy update is received, SPEAR-Q verifies the policy digest, issuer authorization, signature validity, and policy version. The received policy content must match its signed digest, \(I_\pi\) must be the authorized issuer for \(d\), the issuer signature must verify successfully, and the candidate version must be newer than the currently active version. An update that fails any of these checks is rejected before commitment.

Validated updates are then submitted to Raft for ordered commitment across the PAN cluster. An update is committed only after a strict-majority Raft commitment is reached. Raft's role is limited to establishing a consistent commitment order across the distributed cluster, ensuring that different PANs do not treat conflicting update sequences as committed state. It does not re-validate the policy, since authenticity and integrity have already been established.

After receiving a committed update, each PAN applies the policy to its local registry and reports the resulting version and digest. A committed policy becomes authorization-active once a strict majority $Q_P=\lfloor N/2\rfloor+1$ of PANs report the same applied version and digest. This PAN-level activation quorum is distinct from the Raft commitment quorum: Raft establishes consistent commitment of the update, whereas $Q_P$ confirms that the committed state has been applied by a majority of PANs. Once activated, the new version becomes the active policy state for protected data $d$ and is returned for subsequent gateway retrievals.

\begin{algorithm}[t]
\caption{Versioned Sticky Policy Update and Activation}
\label{alg:policy_sync}
\begin{algorithmic}[1]

\Require Signed update
\(\bigl(\pi_d^{new},
M_{\pi}^{new}\bigr)\), where
\(M_{\pi}^{new}=
(d,v_{\pi}^{new},h_{\pi}^{new},t_{\pi},I_{\pi},\sigma_{\pi})\);
current active version \(v_{\pi}^{current}\);
PAN policy registries \(\{\Pi_i\}_{i=1}^{N}\)

\Ensure active policy state
\((v_{\pi}^{new},h_{\pi}^{new})\) or \(\bot\)

\State
\(\widehat{h}_{\pi}^{new}
\gets
\operatorname{SHA256}
(\operatorname{Encode}(\pi_d^{new}))\)

\If{
\(\widehat{h}_{\pi}^{new}\neq h_{\pi}^{new}\)
}
    \State \Return \(\bot\)
\EndIf

\If{\(I_{\pi}\) is not the authorized issuer for \(d\)}
    \State \Return \(\bot\)
\EndIf

\If{
\(\operatorname{Verify}_{pk_{I_{\pi}}}
\bigl(
(d,v_{\pi}^{new},h_{\pi}^{new},t_{\pi}),
\sigma_{\pi}
\bigr)=0
\)
}
    \State \Return \(\bot\)
\EndIf

\If{\(v_{\pi}^{new}\leq v_{\pi}^{current}\)}
    \State \Return \(\bot\)
\EndIf

\State
\(Q_R\gets
\left\lfloor \dfrac{N}{2}\right\rfloor+1\)

\State Submit
\(e_{\pi}\gets
(d,v_{\pi}^{new},h_{\pi}^{new},\pi_d^{new})\)
to Raft

\If{
\(\displaystyle
\sum_{j=1}^{N}
\mathbf{1}
[\operatorname{Ack}_{j}(e_{\pi})=1]
<Q_R
\)
}
    \State \Return \(\bot\)
\EndIf

\State Commit \(e_{\pi}\)

\ForAll{\(PAN_i\) receiving the committed update}
    \If{\(PAN_i\) successfully applies \(e_{\pi}\)}
        \State
        \(\Pi_i[d]\gets
        (v_{\pi}^{new},h_{\pi}^{new})\)
    \EndIf
\EndFor

\State
\(Q_P\gets
\left\lfloor \dfrac{N}{2}\right\rfloor+1\)

\State
\(\mathcal{A}_{\pi}\gets
\left\{
i\in\{1,\ldots,N\}\;:\;
\Pi_i[d]
=
(v_{\pi}^{new},h_{\pi}^{new})
\right\}\)

\If{\(|\mathcal{A}_{\pi}|<Q_P\)}
    \State \Return \(\bot\)
\EndIf

\State
\((v_{\pi}^{current},h_{\pi}^{current})
\gets
(v_{\pi}^{new},h_{\pi}^{new})\)

\State
\Return
\((v_{\pi}^{new},h_{\pi}^{new})\)

\end{algorithmic}
\end{algorithm}

\subsection{Independent PAN Evaluation and Evidence Generation}
\label{sec:evidence_generation}

Policy updates may reach PANs at different times, so a delayed or disconnected PAN can retain an older policy version after a newer version becomes active. Such a PAN is ineligible to contribute authorization evidence for requests governed by the newer version and becomes eligible again once it applies the missing update. More generally, PAN eligibility depends on the policy validation defined in the preceding subsection and additionally requires the PAN's locally applied version and digest to match the active policy state. A PAN that fails either requirement does not contribute evidence for the request. Algorithm \ref{alg:evidence_generation} formalizes the complete PAN evaluation and evidence-generation procedure, including eligibility, input availability, local authorization, and evidence construction.

Each eligible PAN independently evaluates the authorization information available for a request . The local evaluation considers three binary conditions: policy, context, and consent, where a value of 1 indicates that the corresponding authorization requirement is satisfied and a value of 0 indicates that it is not satisfied. These conditions are evaluated independently using the PAN's locally available information.

Authorization also considers a normalized PAN-local user-risk value between 0 and 1, where higher values represent greater risk for the authenticated user. Risk is maintained as PAN-local request-time authorization state and is therefore evaluated during PAN authorization rather than during the policy-update and activation procedure. Risk, rather than a trust score, is used because the authorization rule applies an upper-bound condition: a request remains admissible only while the assessed risk does not exceed the configured threshold \(\rho\). Higher values therefore move the request toward denial, whereas a trust score would require the inverse interpretation. This treatment is consistent with risk-aware authorization in zero-trust architectures, where access decisions are based on current security conditions rather than implicit trust relationships \cite{rose2020zero}. NIST SP 800-207 is used here to motivate continuously evaluated, risk-aware authorization rather than to imply that the standard defines a specific risk-scoring algorithm.

The estimation of the initial risk value from raw behavioural or security information is outside the scope of SPEAR-Q. Instead, the framework assumes that a normalized risk value is supplied by an external risk-assessment mechanism. In the evaluated prototype, controlled initial risk values are provided by the experiment framework to isolate their effect on authorization. The supplied risk value subsequently evolves following committed authorization outcomes, as described in Section~\ref{sec:replication}, consistent with the use of mutable authorization attributes in UCON \cite{park2004ucon}.

Before evaluation, each PAN checks that the required context, consent, and risk information is available. If any of this information is missing, it returns an unavailable result and generates no evidence. Otherwise, the PAN combines the policy, context, consent, and risk conditions using the configured threshold \(\rho=0.60\) to derive a local Permit or Deny decision.

After the local decision is obtained, the PAN constructs a request-bound evidence record containing the request identifier, PAN identity, active policy state, authorization values, local decision, generation time, and a fresh nonce. The resulting evidence record is hashed and digitally signed before being returned for quorum processing. These evidence-construction and signing steps are formalized in Algorithm \ref{alg:evidence_generation}, lines 22–26. Evidence freshness is evaluated later by the Resource Verifier during evidence validation using the evidence-generation time, request time, configured maximum evidence age, and tolerated clock skew. Evidence validation and quorum processing are detailed in Section~\ref{sec:quorum_authorization}.
\begin{algorithm}[t]
\caption{Independent PAN Evidence Generation}
\label{alg:evidence_generation}
\begin{algorithmic}[1]

\Require Authenticated request
\(X=(rid,role,d,a,t,\ell)\) with session-bound identity \(u\);
sticky policy \(\pi_d\) with metadata
\(M_{\pi}=(d,v_{\pi},h_{\pi},t_{\pi},I_{\pi},\sigma_{\pi})\);
local information
\(IB_i=(\Pi_i,\mathcal{C}_i,\mathcal{CO}_i,\mathcal{R}_i)\);
private key \(sk_i\)

\Ensure Signed evidence \((E_i,\sigma_i)\) or \(\bot\)

\If{\(X\) is invalid or \(\pi_d\) is unavailable}
    \State \Return \(\bot\)
\EndIf

\State
\(\widehat{h}_{\pi}\gets
\operatorname{SHA256}
(\operatorname{Encode}(\pi_d))\)

\If{\(I_{\pi}\) is not the authorized issuer for \(d\)}
    \State \Return \(\bot\)
\EndIf

\If{
\(\widehat{h}_{\pi}\neq h_{\pi}\)
\(\lor\)
\(\operatorname{Verify}_{pk_{I_{\pi}}}
((d,v_{\pi},h_{\pi},t_{\pi}),\sigma_{\pi})=0\)
}
    \State \Return \(\bot\)
\EndIf

\If{
\(\Pi_i[d]\neq(v_{\pi},h_{\pi})\)
}
    \State \Return \(\bot\)
\EndIf

\If{required context, consent, or risk information is unavailable in \(IB_i\)}
    \State \Return \(\bot\)
\EndIf

\State
\(P_i\gets
\operatorname{EvalPolicy}(\pi_d,X)\)

\State
\(C_i\gets
\operatorname{EvalContext}(X,\mathcal{C}_i)\)

\State
\(Co_i\gets
\operatorname{EvalConsent}(u,d,a,\mathcal{CO}_i)\)

\State
Read \(R_i^{(k)}\) from \(\mathcal{R}_i\)

\State
\(D_i\gets
\mathbf{1}
\left[
P_i=C_i=Co_i=1
\;\land\;
R_i^{(k)}\leq\rho
\right]\),
\(\rho=0.60\)

\State Generate fresh nonce \(n_i\) and evidence time \(t_i\)

\State
\(E_i\gets
(rid,i,v_{\pi},h_{\pi},
P_i,C_i,Co_i,R_i^{(k)},D_i,t_i,n_i)\)

\State
\(h_{E_i}\gets
\operatorname{SHA256}
(\operatorname{Encode}(E_i))\)

\State
\(\sigma_i\gets
\operatorname{Sign}_{sk_i}(h_{E_i})\)

\State \Return \((E_i,\sigma_i)\)

\end{algorithmic}
\end{algorithm}

\subsection{Evidence Quorum Authorization}
\label{sec:quorum_authorization}

A single PAN decision does not determine global authorization. SPEAR-Q therefore uses the strict-majority Evidence quorum \(Q_E=\lfloor N/2\rfloor+1\). Thus, for the evaluated PAN Cluster sizes,
\(N=3\) gives \(Q_E=\lfloor 3/2\rfloor+1=2\),
\(N=5\) gives \(Q_E=\lfloor 5/2\rfloor+1=3\), and
\(N=7\) gives \(Q_E=\lfloor 7/2\rfloor+1=4\).
The corresponding numbers of unavailable PANs that can be tolerated while still retaining quorum are obtained from \(N-Q_E\), giving \(1\), \(2\), and \(3\) unavailable PANs for \(N=3\), \(5\), and \(7\), respectively. This bound concerns PAN-quorum formation rather than complete system
availability. For request \(X\), the Resource Verifier admits or discards each received PAN evidence record through the sequence of checks described below. Evidence that fails validation or is otherwise inadmissible is excluded from quorum computation and is not treated as a Deny vote. Global authorization is derived only after evidence validation is complete: a Permit is issued when at least \(Q_E\) distinct admissible records support Permit, regardless of additional Deny-supporting or excluded records. If fewer than \(Q_E\) admissible Permit-supporting records are available, SPEAR-Q returns Deny under its fail-closed authorization semantics. SPEAR-Q therefore retains a binary Permit/Deny request-time outcome, while excluded or anomalous evidence can be retained for audit and offline diagnosis.

The Resource Verifier first validates the request's active sticky policy using the policy-validation procedure defined in the preceding subsection. For each received evidence record, it then checks that the record is well formed, bound to the current request, and carries the active policy version and digest. Records with a PAN identity or nonce already admitted for the request, or that fail the freshness requirement, are discarded. Surviving records then undergo PAN-signature verification. The verifier extracts the signed policy, context, consent, and risk values and reconstructs the local decision; evidence is admitted only when the reconstructed decision matches the decision reported by the PAN.

This subsection defines evidence validation and strict-majority Evidence-quorum derivation. The resulting global decision and verified evidence are subsequently used to construct and commit the authorization record and, where applicable, update PAN-local risk state, as described in Section~\ref{sec:replication}.

\subsection{Authorization Commitment and PAN-Local Risk Evolution}
\label{sec:replication}

After the Evidence quorum checks described in Section~\ref{sec:quorum_authorization} are completed, the resulting global decision and verified evidence are used to construct an authorization record. This record binds the request identifier, global decision, verified evidence, and active policy state, and is then submitted to Raft for commitment. The record is committed only when a strict majority of acknowledgments is reached; if quorum is not reached, the procedure returns the global decision and verified evidence without committing, and no resource release or risk update follows (lines 26 to 29). Raft records the resulting authorization state but does not generate PAN evidence, participate in the Evidence quorum, or alter the global decision \(D_g\).

For a committed global Permit, the Resource Verifier issues an authorization certificate binding \(rid\), protected data, requested action, active policy, global decision, and verified PAN identities. The protected-data provider
validates this certificate before releasing the resource. A global Deny results in no resource release.

A successfully committed authorization record also updates the local risk state of each PAN that contributed admitted evidence. Each contributing PAN updates its local risk value once according to its own local decision: a local Permit decreases the value by 0.05, bounded below by zero, while a local Deny increases it by 0.10, bounded above by one (lines 30 to 33). These fixed update values are prototype configuration parameters used to evaluate directional PAN-local risk evolution rather than a calibrated behavioural-risk model; estimation of the underlying risk value from raw behavioural or security information remains outside the scope of SPEAR-Q, as noted in Section~\ref{sec:evidence_generation}. Uncommitted, failed, duplicated, or replayed records cause no risk update. Each PAN retains the identifiers of previously applied requests so that a committed request can update its local risk state at most once. Because each PAN applies the update using its own local decision, PAN risk values can diverge over time: for the same committed request, a PAN returning a local Deny can increase its risk value while another PAN returning a local Permit decreases its value. These independently maintained risk states are used in subsequent PAN evaluations.

The authorization certificate applies only to the request for which it was issued and does not grant continuing access to the protected object. Because the sticky policy remains associated with the protected data, each subsequent access or sharing request is treated as a new authorization request with a distinct request identifier and independently generated PAN evidence. The resulting authorization outcome is committed as a separate authorization record binding the request, verified evidence, decision, and active policy state, thereby providing a request-level record of policy enforcement for subsequent audit.

Together, Sections~\ref{sec:quorum_authorization} and \ref{sec:replication} cover the complete procedure formalized in Algorithm \ref{alg:quorum_authorization}: request-policy validation and evidence validation (lines 1 to 24), Raft-based commitment of the authorization record (lines 25 to 29), PAN-local risk evolution (lines 30 to 33), and conditional certificate issuance and resource release (lines 34 to 38).

\begin{algorithm}[!t]
\caption{Evidence Quorum Authorization and Commitment}
\label{alg:quorum_authorization}
\begin{algorithmic}[1]

\Require \(X=(rid,role,d,a,t,\ell)\), \(\pi_d\),
\(\mathcal{M}_X\), PAN public keys, \(N\)
\Ensure \(D_g\), verified evidence \(\mathcal{V}_X\)

\State \(Q_E,Q_R\gets\lfloor N/2\rfloor+1\);
\(\mathcal{V}_X,\mathcal{I}_X,\mathcal{N}_X\gets\emptyset\)

\State \(\hat h_{\pi}\gets
\operatorname{SHA256}(\operatorname{Encode}(\pi_d))\)

\If{\(X\) is invalid \(\lor\)
\(\hat h_{\pi}\neq h_{\pi}\) \(\lor\)
\(I_{\pi}\) is unauthorized \(\lor\)
\(\operatorname{Verify}_{pk_{I_{\pi}}}
((d,v_{\pi},h_{\pi},t_{\pi}),\sigma_{\pi})=0\)}
    \State \Return \(\textsc{Deny},\mathcal{V}_X\)
\EndIf

\ForAll{\((E_i,\sigma_i)\in\mathcal{M}_X\)}

    \If{\(E_i\) is malformed}
        \State \textbf{continue}
    \EndIf

    \State Extract
    \(rid_i,i,v_i,h_i,P_i,C_i,Co_i,R_i^{(k)},D_i,t_i,n_i\)

    \If{\(rid_i\neq rid\) \(\lor\)
    \(v_i\neq v_{\pi}\) \(\lor\) \(h_i\neq h_{\pi}\) \(\lor\)
    \(i\in\mathcal{I}_X\) \(\lor\) \(n_i\in\mathcal{N}_X\) \(\lor\)
    \(|t_i-t|>\Delta_E+\epsilon\)}
        \State \textbf{continue}
    \EndIf

    \State \(\hat h_{E_i}\gets
    \operatorname{SHA256}(\operatorname{Encode}(E_i))\)

    \If{\(\operatorname{Verify}_{pk_i}(\hat h_{E_i},\sigma_i)=0\)}
        \State \textbf{continue}
    \EndIf

    \State
    \(\hat D_i\gets
    \mathbf{1}[P_i=C_i=Co_i=1
    \land R_i^{(k)}\leq\rho]\), \(\rho=0.60\)

    \If{\(\hat D_i\neq D_i\)}
        \State \textbf{continue}
    \EndIf

    \State
    \(\mathcal{V}_X\gets\mathcal{V}_X\cup\{(E_i,\sigma_i)\}\);
    \(\mathcal{I}_X\gets\mathcal{I}_X\cup\{i\}\);
    \(\mathcal{N}_X\gets\mathcal{N}_X\cup\{n_i\}\)

\EndFor

\State
\(D_g\gets
\mathbf{1}\!\left[
|\mathcal{I}_X|\geq Q_E
\land
\sum_{i\in\mathcal{I}_X}D_i\geq Q_E
\right]\)

\State
\(\mathcal{R}_X\gets
(rid,D_g,\mathcal{V}_X,v_{\pi},h_{\pi})\)

\If{\(\sum_{j=1}^{N}
\mathbf{1}[\operatorname{Ack}_j(\mathcal{R}_X)=1]<Q_R\)}
    \State \Return \(D_g,\mathcal{V}_X\)
\EndIf

\State Commit \(\mathcal{R}_X\)

\ForAll{\(i\in\mathcal{I}_X\) not previously updated for \(rid\)}
    \State
    \(R_i^{(k+1)}\gets
    D_i\max(0,R_i^{(k)}-0.05)
    +(1-D_i)\min(1,R_i^{(k)}+0.10)\)
    \State Mark \(rid\) as applied at \(PAN_i\)
\EndFor

\If{\(D_g=1\)}
    \State Issue \(\mathcal{A}_X\) bound to
    \((rid,d,a,\pi_d,D_g,\mathcal{I}_X)\)
    \State Release \(d\) only if the provider validates \(\mathcal{A}_X\)
\EndIf

\State \Return \(D_g,\mathcal{V}_X\)

\end{algorithmic}
\end{algorithm}

\raggedbottom
\section{Evaluation and Results}
\label{sec:experimentation}

To evaluate SPEAR-Q under distributed execution, we deployed the complete
prototype across physically separated hosts in the Airbus CyberSecurity
Simulation Platform \footnote{https://www.cyber.airbus.com/en/products/cyberrange}. Measurements were collected at both the request level and at internal processing boundaries, allowing the client-observed
authorization cost to be related to the individual stages that produce it. The same deployment was subsequently used for controlled policy, evidence, quorum, and post-authorization experiments so that performance and authorization behaviour were evaluated under a common implementation. This also allows us to evaluate how sticky policy activation and request-time authorization are distributed across the PAN cluster rather than controlled by a central PDP.

\subsection{Implementation and Experimental Setup}
\label{subsec:experimental_setup}

SPEAR-Q was implemented in Python 3.14.1 as independent services for
three active-active gateways, PANs, the Resource Verifier,
protected-data provider, controller, and Raft coordination layer.
Flask provides the HTTP service interfaces, while \texttt{requests} and
\texttt{aiohttp} support service communication and concurrent workload
generation. PAN evidence is deterministically encoded, hashed with
SHA-256, and digitally signed and verified using the Python
\texttt{cryptography} library. \texttt{PyYAML}, \texttt{psutil}, and
NumPy support configuration, service-health checks, and result
aggregation, respectively.

The same code and configuration were deployed on all hosts.
Before each independent execution, an automated preflight check verified
the participating services, active cluster configuration, and clock
alignment required for evidence freshness validation. Between repetitions, gateway replay records, PAN-local request state, verifier state, and experimental state were reset to ensure that each run started from an independent state.

\begin{figure*}[!t]
\centering
\includegraphics[width=0.92\textwidth]
{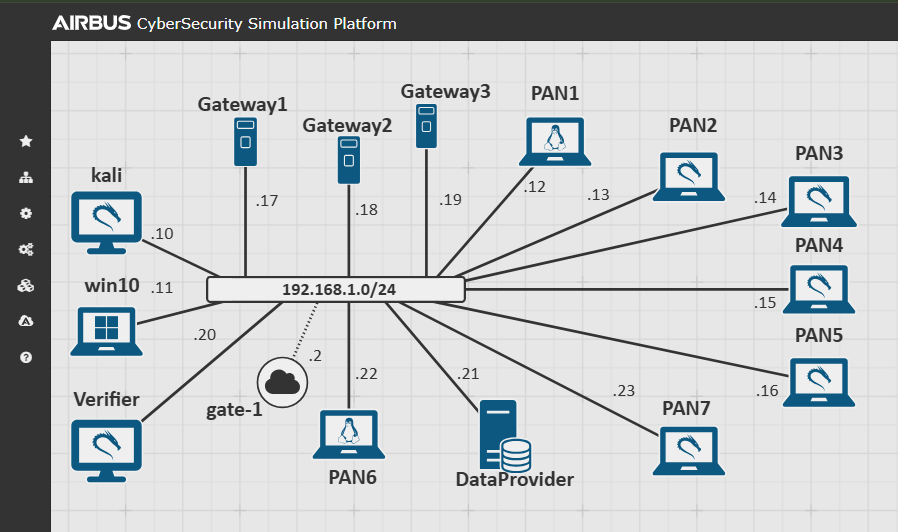}
\caption{Multi-host SPEAR-Q deployment in the Airbus CyberSecurity
Simulation Platform. Seven PANs, three active-active gateways, the
Resource Verifier, controller, client, and protected-data provider
communicate over the isolated experimental subnet.}
\label{fig:airbus_normal_topology}
\end{figure*}

The physical deployment comprised seven PAN hosts, three gateways, a
controller, a Resource Verifier, a Windows client, and a Windows
protected-data provider on the \texttt{192.168.1.0/24} experimental
subnet. Figure~\ref{fig:airbus_normal_topology} shows the deployment,
while Table~\ref{tab:airbus_hosts} records the host assignment.

\begin{table}[!t]
\caption{Airbus cyber-range hosts used in the evaluation.}
\label{tab:airbus_hosts}
\centering
\footnotesize
\setlength{\tabcolsep}{3pt}
\renewcommand{\arraystretch}{1.08}

\begin{tabular}{|l|l|l|l|}
\hline
\textbf{Host} & \textbf{Address} & \textbf{Platform} & \textbf{Role} \\
\hline
KALI     & 192.168.1.10 & Linux   & Controller \\
\hline
WIN10    & 192.168.1.11 & Windows & Client \\
\hline
PAN1     & 192.168.1.12 & Linux   & PAN \\
\hline
PAN2     & 192.168.1.13 & Linux   & PAN \\
\hline
PAN3     & 192.168.1.14 & Linux   & PAN \\
\hline
PAN4     & 192.168.1.15 & Linux   & PAN \\
\hline
PAN5     & 192.168.1.16 & Linux   & PAN \\
\hline
GW1      & 192.168.1.17 & Linux   & Gateway \\
\hline
GW2      & 192.168.1.18 & Linux   & Gateway \\
\hline
GW3      & 192.168.1.19 & Linux   & Gateway \\
\hline
VERIFIER & 192.168.1.20 & Linux   & Verifier \\
\hline
PROVIDER & 192.168.1.21 & Windows & Provider \\
\hline
PAN6     & 192.168.1.22 & Linux   & PAN \\
\hline
PAN7     & 192.168.1.23 & Linux   & PAN \\
\hline
\end{tabular}

\end{table}

The evaluation used a deterministic healthcare workload derived from MIMIC-IV clinical records. MIMIC-IV was accessed under the applicable PhysioNet Data Use Agreement,\footnote{MIMIC-IV is available through PhysioNet at \url{https://physionet.org/content/mimiciv/3.1/}} and no credentialed source records were redistributed. MIMIC-IV supplied representative protected healthcare objects, while the experiment scripts assigned the role, action, context, consent, location, credential, and PAN-local risk
conditions required for each test case. Scenario labels were not provided
to the gateways, PANs, or Resource Verifier; the deployed services
processed only the authorization request and its associated policy state.

The resulting pool contained 40,000 requests, as summarized in
Table~\ref{tab:workload}. The workload pool defines the source set of request scenarios and is distinct from the number of request executions performed in the performance evaluation.Performance experiments used legitimate requests to measure latency and throughput over a consistent successful authorization path, while authorization-violation and policy-attack requests were evaluated separately in the mechanism experiments because they exercise different rejection and fail-closed paths. Comparable runs used the same request set and the fixed seed \texttt{20260820}.

\begin{table}[!t]
\caption{Composition of the deterministic healthcare workload.}
\label{tab:workload}
\centering
\small
\setlength{\tabcolsep}{5pt}
\renewcommand{\arraystretch}{1.08}

\begin{tabular}{|l|r|r|}
\hline
\textbf{Request class} & \textbf{Count} & \textbf{Share} \\
\hline
Legitimate requests      & 20,000 & 50\% \\
\hline
Authorization violations & 14,000 & 35\% \\
\hline
Protocol/policy attacks  & 6,000 & 15\% \\
\hline
\textbf{Total}           & \textbf{40,000} & \textbf{100\%} \\
\hline
\end{tabular}

\end{table}

For the performance evaluation, the PAN cluster size was varied over
$N\in\{3,5,7\}$. The corresponding evidence quorum was the strict
majority, $Q_E=\lfloor N/2 \rfloor+1$, giving $Q_E\in\{2,3,4\}$. Request concurrency, $C$, denoting the number of simultaneously active
requests, was varied over $C\in\{1,10,25,50,75,100\}$. Each configuration processed 100 legitimate requests in five independent repetitions, yielding 500 request executions per configuration. The resulting
performance study therefore comprised 18 configurations, 90 independent experimental runs (18 configurations $\times$ 5 repetitions), and 9,000 request executions in total (90 runs $\times$ 100 requests).

\begin{figure*}[!t]
\centering

\begin{minipage}{0.49\textwidth}
\centering
\includegraphics[width=\linewidth]{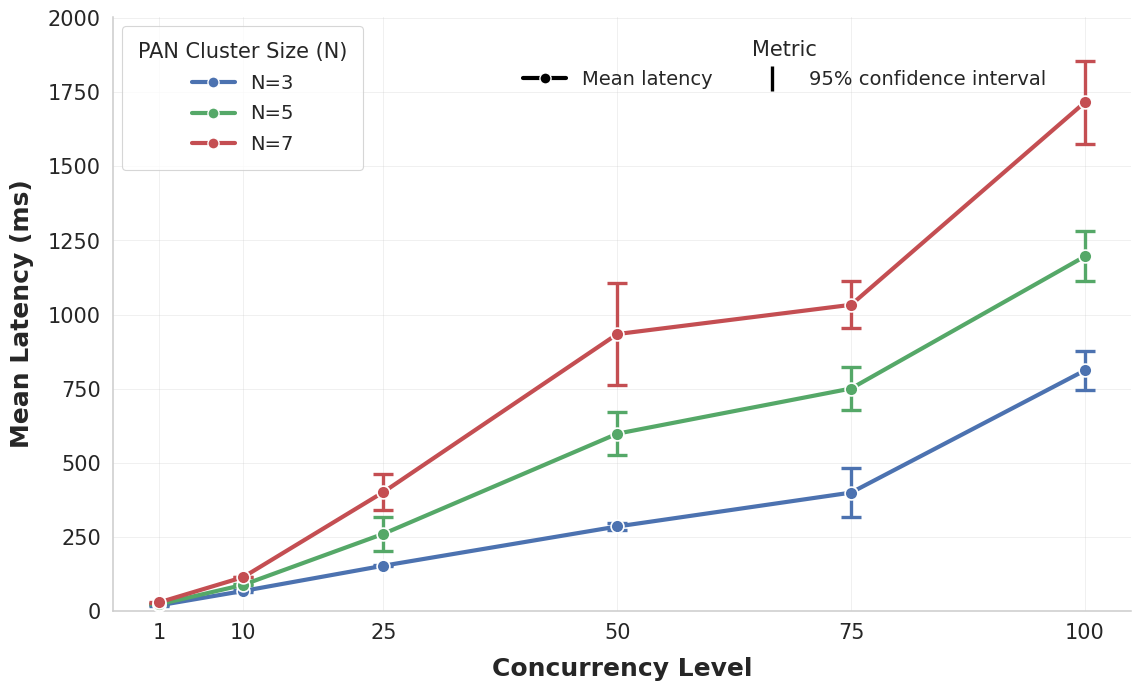}
\\[1mm]
\textbf{(a)} End-to-end latency
\end{minipage}
\hfill
\begin{minipage}{0.49\textwidth}
\centering
\includegraphics[width=\linewidth]{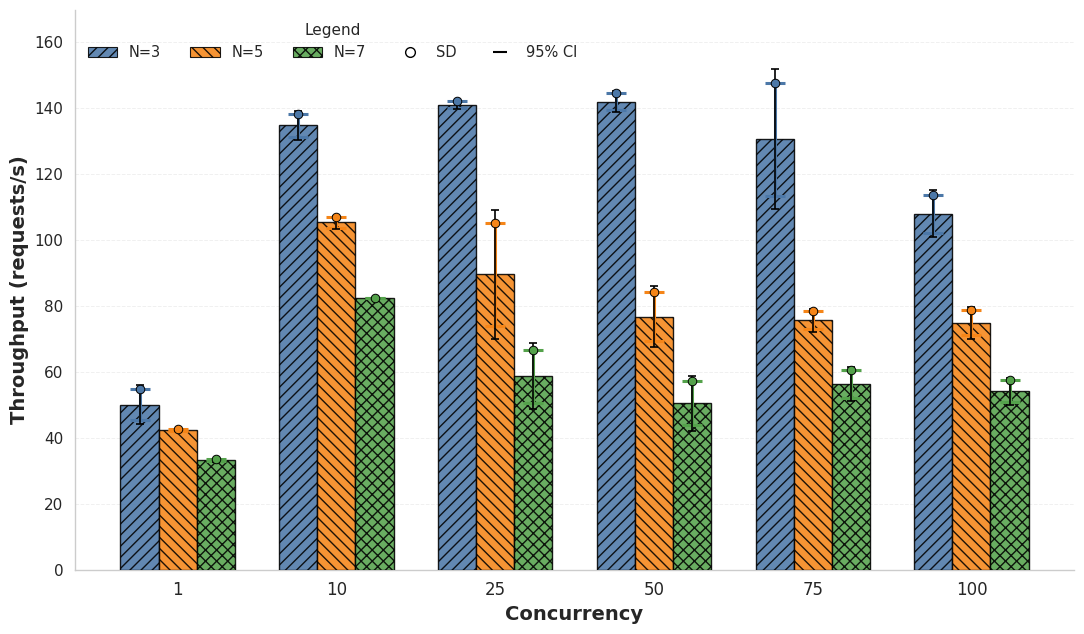}
\\[1mm]
\textbf{(b)} Completed throughput
\end{minipage}

\caption{End-to-end SPEAR-Q performance across PAN cluster sizes
$N\in\{3,5,7\}$ and concurrency levels
$C\in\{1,10,25,50,75,100\}$.}
\label{fig:performance_scaling}
\end{figure*}

The number of repetitions was determined before the main performance evaluation using variability observed in pilot runs and a target relative 95\% confidence-interval half-width of 5\%. For each run-level metric, a two-sided 95\% confidence interval for the mean was calculated using Student's \(t\)-distribution as
\(\bar{x} \pm t_{0.975,r-1}s/\sqrt{r}\),
where \(\bar{x}\) is the mean across the repetitions, \(s\) is the corresponding standard deviation, \(r\) is the number of repetitions, and \(t_{0.975,r-1}\) is the corresponding critical value from the \(t\)-distribution. Each repetition represented a complete independent execution of the system rather than an individual request within a longer run, because requests processed within the same execution share system state and therefore cannot be treated as independent measurements of run-to-run variability.


\subsection{Performance and Authorization-Cost Results}
\label{subsec:multihost_results}

This subsection evaluates the performance of the deployed SPEAR-Q authorization path and identifies where its processing cost is incurred. End-to-end scaling with PAN cluster size and concurrency is examined first, followed by a decomposition of authorization latency and PAN-side processing cost.

\subsubsection{End-to-End Performance and Saturation}

The end-to-end evaluation first examines how PAN cluster size and concurrent request load affect authorization latency and throughput. Figure~\ref{fig:performance_scaling} reports the end-to-end behaviour as
both the PAN cluster size and concurrent request load increase. Concurrency
denotes the maximum number of requests simultaneously in flight; each repetition contains 100 requests. Consequently, latency measures the
response time experienced by an individual request, whereas throughput
measures aggregate completions over wall-clock time.

At $C=1$, mean latency is 19.98~ms for $N=3$, 23.48~ms for $N=5$,
and 29.97~ms for $N=7$, with corresponding p99 values of 25.43,
28.68, and 34.41~ms. The lightly loaded measurements therefore show a
moderate increase as the PAN cluster grows. Under concurrent load, this
separation becomes larger: at $C=25$, mean latency reaches 153.67,
261.24, and 401.96~ms, and at $C=100$ it reaches 812.07, 1196.30,
and 1715.64~ms for $N=3,5,7$, respectively. The corresponding p99
values at $C=100$ are 925.73, 1334.06, and 1847.92~ms.
Figure~\ref{fig:performance_scaling}(a) therefore shows that authorization
latency increases with both cluster size and concurrency, with the effect
of larger clusters becoming more pronounced under high load.

Figure~\ref{fig:performance_scaling}(b) shows how concurrency affects
completed throughput. At $C=1$, throughput is approximately the reciprocal
of service time: the 19.98~ms mean for $N=3$ corresponds to approximately
$1000/19.98=50.1$ requests/s, closely matching the observed
50.08 requests/s, with the same consistency observed for $N=5$ and $N=7$.
For $N=3$, throughput rises to 134.61 requests/s at $C=10$ and then remains close to this level at $C=25$ and $C=50$, reaching 140.87 and 141.81 requests/s, respectively. This plateau indicates that increasing the number of concurrently active requests no longer produces a proportional increase in completed work. At higher concurrency, multiple requests simultaneously traverse the distributed authorization path, increasing the demand on PAN-local processing and distributed evidence collection, and the resulting high-load cost, shown in the latency decomposition in Figure~\ref{fig:latency_breakdown}, is concentrated primarily in PAN evidence waiting and PAN-local risk-update processing, with Evidence-quorum calculation contributing only a negligible fraction of the authorization path.

At C=100, throughput falls to 107.85 requests per second as PAN evidence waiting and risk-update processing, the dominant cost drivers identified above, accumulate additional delay under concurrent load. Larger clusters reach this saturation point even sooner, with N=5 and N=7 achieving their highest measured throughput as early as C=10, at 105.24 and 82.44 requests per second respectively, because more PANs participating in evidence generation and a larger strict-majority Evidence quorum requirement compound the same underlying bottleneck. Overall, Figure~\ref{fig:performance_scaling}(b) shows that concurrency improves throughput only up to the saturation region, beyond which additional in-flight requests no longer increase completed throughput proportionally.

Table~\ref{tab:performance_summary} condenses the low-load, peak-throughput,
and high-load operating points. Run-to-run mean-latency variability also increases under heavy load. For example, the mean-latency standard
deviation for $N=7$ increases from 0.28~ms at $C=1$ to 112.74~ms at
$C=100$. All 9,000 performance requests completed without execution failure or timeout, showing that the observed high-load behaviour reflects saturation rather than loss of request completion.

\begin{table}[!t]
\caption{Representative performance points across five independent
repetitions. Values report run-level means.}
\label{tab:performance_summary}
\centering
\scriptsize
\setlength{\tabcolsep}{3.2pt}
\renewcommand{\arraystretch}{1.15}

\begin{tabular}{|c|c|c|c|c|}
\hline
\textbf{$N$} &
\shortstack{\textbf{Mean latency}\\\textbf{$C=1$ (ms)}} &
\shortstack{\textbf{Peak throughput}\\\textbf{(req/s)}} &
\shortstack{\textbf{Mean latency}\\\textbf{$C=100$ (ms)}} &
\shortstack{\textbf{p99}\\\textbf{$C=100$ (ms)}} \\
\hline

3 &
19.98 &
141.81 ($C=50$) &
812.07 &
925.73 \\
\hline

5 &
23.48 &
105.24 ($C=10$) &
1196.30 &
1334.06 \\
\hline

7 &
29.97 &
82.44 ($C=10$) &
1715.64 &
1847.92 \\
\hline

\end{tabular}

\end{table}

\subsubsection{Authorization-Path Latency Decomposition}

\begin{figure*}[!t]
\centering

\begin{minipage}{0.49\textwidth}
\centering
\includegraphics[width=\linewidth]{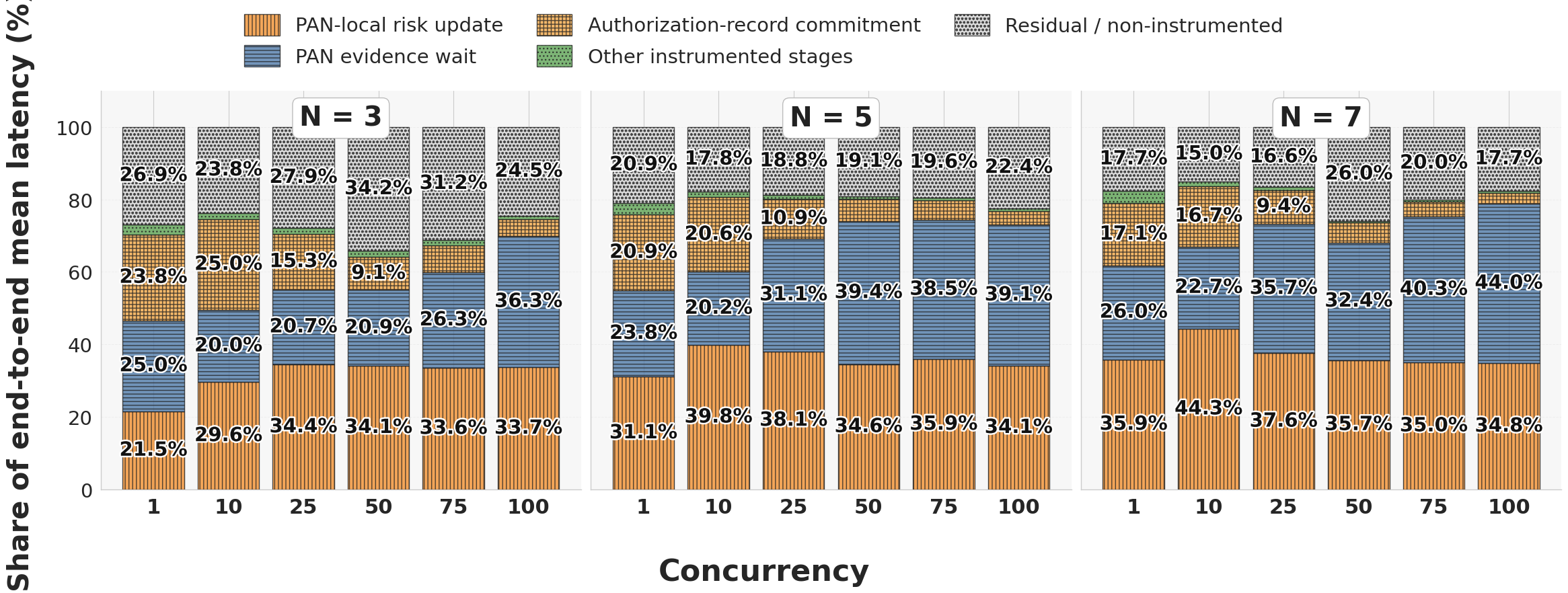}
\\[1mm]
\textbf{(a)} Relative stage contribution
\end{minipage}
\hfill
\begin{minipage}{0.49\textwidth}
\centering
\includegraphics[width=\linewidth]{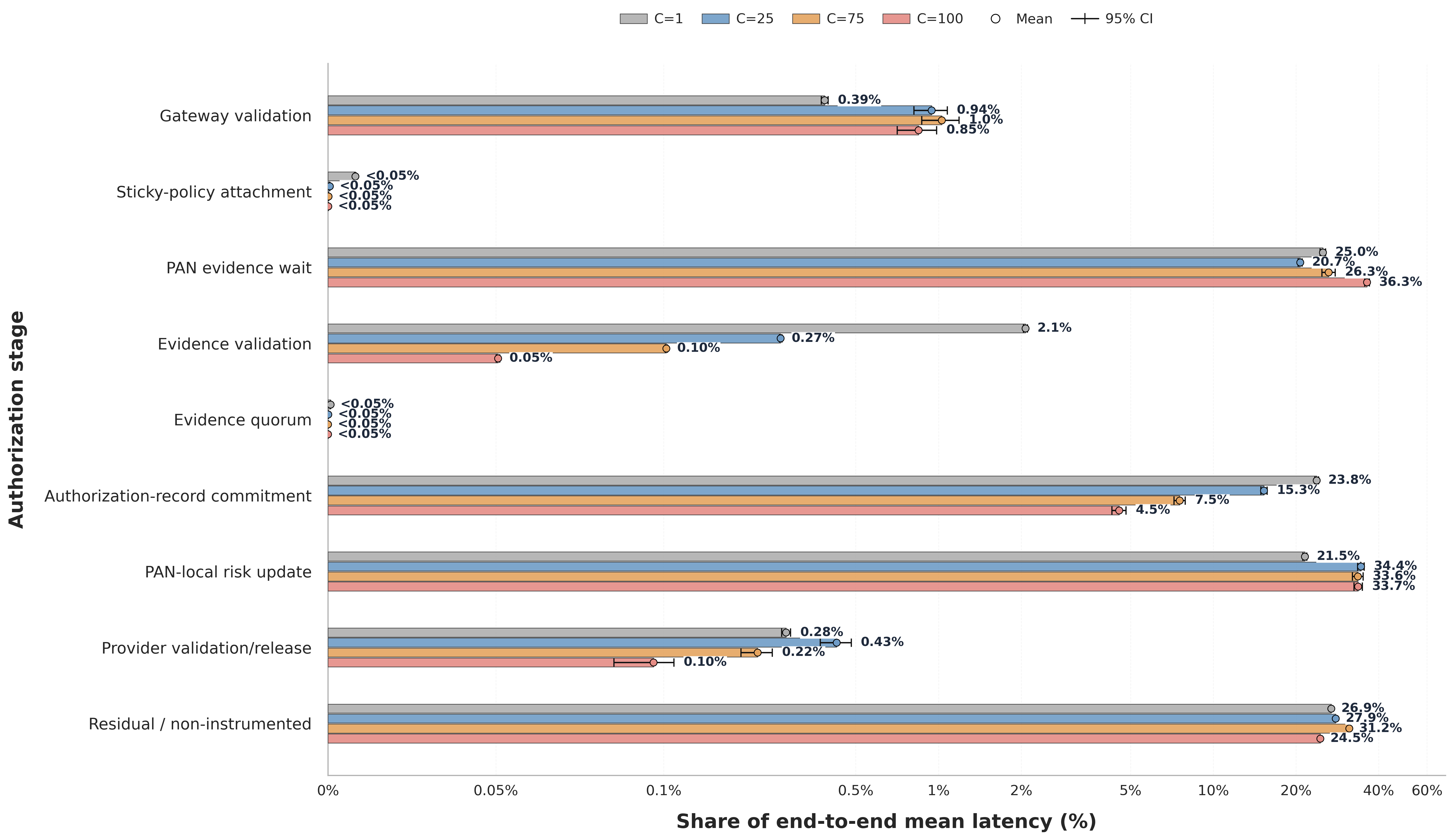}
\\[1mm]
\textbf{(b)} Detailed relative stage contribution
\end{minipage}

\caption{Authorization-path latency decomposition. Panel (a) shows the relative contribution of the main latency components across $N\in\{3,5,7\}$ and $C\in\{1,10,25,50,75,100\}$. Panel (b) provides a detailed relative stage breakdown for $N=3$ at $C\in\{1,25,75,100\}$. All contributions are expressed as percentages of the corresponding end-to-end mean latency. The residual component represents latency outside the explicitly instrumented stage boundaries. PAN evidence wait denotes the parallel critical-path interval required to collect sufficient admissible evidence rather than the sum of processing times across PANs.}
\label{fig:latency_breakdown}
\end{figure*}

The end-to-end results are next decomposed to identify which stages contribute most to the observed authorization latency. End-to-end measurements reveal the onset of saturation but do not indicate which authorization stages contribute to the associated latency increase. The authorization path was therefore instrumented at gateway validation, sticky policy attachment, PAN evidence wait, evidence validation, Evidence-quorum calculation, authorization-record commitment, PAN-local risk update, and provider validation/release. Figure~\ref{fig:latency_breakdown} reports the resulting decomposition.

Figure \ref{fig:latency_breakdown}(a) compares the end-to-end latency composition across all evaluated cluster sizes N = {3, 5, 7} and concurrency levels C = {1, 10, 25, 50, 75, 100}.  The relative contribution of the stages changes with both cluster size and concurrency. For N = 7, PAN evidence waiting increases from 26.0\% of end-to-end latency at C = 1 to 44.0\% at C = 100, while PAN-local risk update changes from 35.9\% to 34.8\%. Over the same range, authorization-record commitment decreases from 17.1\% to 2.9\%. The remaining directly instrumented stages contribute comparatively small shares, with the residual component accounting for latency outside the explicitly instrumented boundaries.

Figure \ref{fig:latency_breakdown}(b) fixes the cluster at N = 3 to isolate the effect of concurrency from the additional coordination introduced by larger PAN clusters; the cross-cluster behaviour remains visible in Figure 6(a). At C = 1, PAN evidence waiting, authorization-record commitment, and PAN-local risk update contribute 25.0\%, 23.8\%, and 21.5\%, respectively. At C = 25, PAN-local risk update becomes the largest instrumented stage at 34.4\%, compared with 20.7\% for PAN evidence waiting and 15.3\% for authorization-record commitment. At C = 75, the corresponding contributions are 33.6\%, 26.3\%, and 7.5\%. At C = 100, PAN evidence waiting increases to 36.3\% and exceeds PAN-local risk update at 33.7\%, while authorization-record commitment falls to 4.5\%. Gateway validation remains at or below 1\%, and sticky policy attachment and Evidence-quorum calculation remain negligible across these detailed configurations.

The cross-configuration view in Figure \ref{fig:latency_breakdown}(a) and the fixed-cluster stage view in Figure \ref{fig:latency_breakdown}(b) are consistent with the authorization mechanisms defined in Sections \ref{sec:evidence_generation} to Section \ref{sec:replication}. PAN evidence waiting corresponds to the distributed execution path initiated in Section \ref{sec:evidence_generation} and Algorithm \ref{alg:evidence_generation}: each eligible PAN must validate the active policy state, evaluate the policy, context, consent, and risk conditions, construct the evidence record, and sign it before the Resource Verifier can collect sufficient admissible evidence for authorization. The verifier therefore waits on distributed PAN execution and communication rather than performing a purely local operation. As concurrency increases, contention and scheduling across these independently executing services make this interval increasingly visible in the end-to-end path. Nevertheless, the verifier cannot derive a decision until enough admissible PAN evidence has been collected, making evidence generation, transmission, and collection part of the critical authorization path.

PAN-local risk update maps to the post-commitment state evolution defined in Section \ref{sec:replication} and Algorithm \ref{alg:quorum_authorization}. Although the numerical risk adjustment itself is simple, the update is applied independently to each PAN that contributed admitted evidence and also records the request as applied. Its measured cost therefore reflects distributed PAN-local state processing rather than the arithmetic operation alone. By contrast, the Evidence-quorum calculation in Section \ref{sec:quorum_authorization} and Algorithm \ref{alg:quorum_authorization} is performed after evidence validation and reduces to checking the number of distinct admitted records and Permit decisions against the strict-majority threshold. With only N = 3, 5, or 7 PANs in the evaluated clusters, this local threshold calculation is expected to be substantially cheaper than waiting for distributed evidence generation or completing PAN-local state updates. The observed negligible quorum-calculation share is therefore consistent with the algorithmic structure rather than being an unexpected measurement result.

Sticky policy attachment at the gateway remains below 1\% across the evaluated configurations. This measurement corresponds only to policy retrieval and forwarding and should not be interpreted as the complete cost of sticky policy enforcement, since policy validation and request evaluation occur subsequently during PAN processing

\subsubsection{PAN-Side Processing Cost}

\begin{figure}[!t]
\centering
\includegraphics[width=0.98\columnwidth]{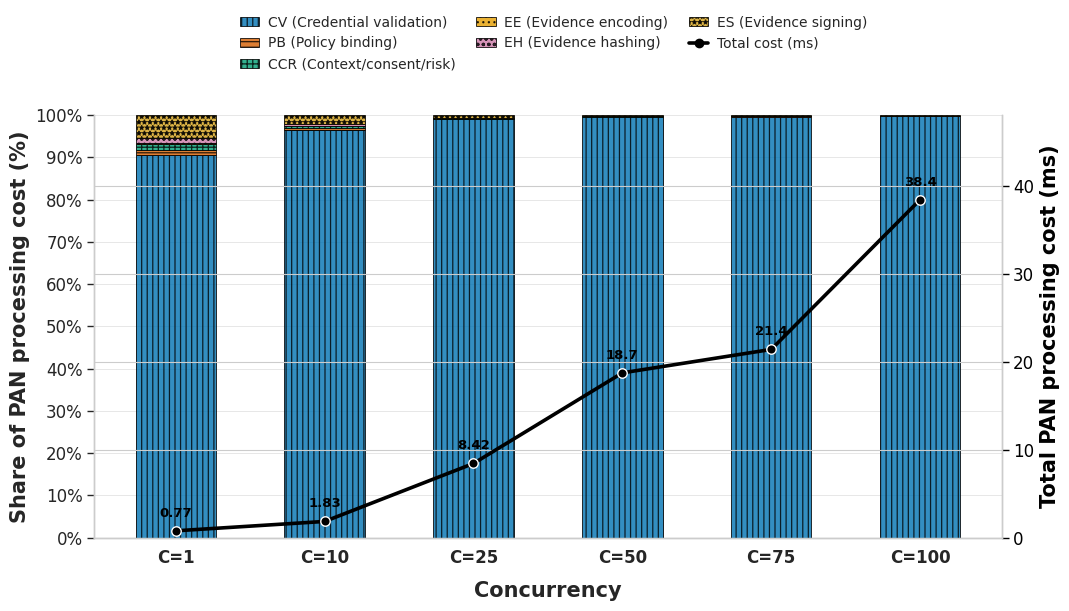}
\caption{PAN-side processing cost for $N=7$ across concurrency levels. Stacked bars show the relative contribution of each PAN operation, while the overlaid line reports total mean PAN processing cost.}
\label{fig:pan_processing_cost}
\end{figure}

\begin{table}[!t]
\caption{PAN-side processing-cost composition for $N=7$.
Component values are percentages of total PAN processing cost;
the final row reports the corresponding total mean cost in milliseconds.}
\label{tab:pan_processing_breakdown}
\centering
\scriptsize
\setlength{\tabcolsep}{3.2pt}
\renewcommand{\arraystretch}{1.14}

\begin{tabular}{|l|r|r|r|r|r|r|}
\hline
\textbf{PAN operation} &
\textbf{$C=1$} &
\textbf{$C=10$} &
\textbf{$C=25$} &
\textbf{$C=50$} &
\textbf{$C=75$} &
\textbf{$C=100$} \\
\hline

Credential validation (CV)
& 90.48\%
& 96.39\%
& 99.06\%
& 99.60\%
& 99.64\%
& 99.78\% \\
\hline

Policy binding (PB)
& 1.37\%
& 0.54\%
& 0.13\%
& 0.06\%
& 0.05\%
& 0.03\% \\
\hline

Context/consent/risk (CCR)
& 1.37\%
& 0.46\%
& 0.11\%
& 0.06\%
& 0.05\%
& 0.02\% \\
\hline

Evidence encoding (EE)
& 0.25\%
& 0.07\%
& 0.02\%
& 0.01\%
& 0.01\%
& 0.00\% \\
\hline

Evidence hashing (EH)
& 1.17\%
& 0.46\%
& 0.10\%
& 0.05\%
& 0.04\%
& 0.04\% \\
\hline

Evidence signing (ES)
& 5.37\%
& 2.07\%
& 0.59\%
& 0.22\%
& 0.22\%
& 0.13\% \\
\hline

\textbf{Total cost (ms)}
& \textbf{0.768}
& \textbf{1.834}
& \textbf{8.422}
& \textbf{18.718}
& \textbf{21.382}
& \textbf{38.400} \\
\hline

\end{tabular}
\end{table}

The analysis then examines PAN-side processing in greater detail to identify the local operations that dominate this component of the authorization path. Figure~\ref{fig:pan_processing_cost} and Table~\ref{tab:pan_processing_breakdown} summarize the relative PAN-side processing profile and the corresponding numerical breakdown for the \(N=7\) configuration across concurrency levels. Total PAN processing increases from 0.768~ms at $C=1$ to 8.422~ms at
$C=25$ and 38.400~ms at $C=100$. Credential validation accounts for
90.48\% of PAN cost at $C=1$ and increases to 99.78\% at $C=100$.
In contrast, evidence signing falls from 5.37\% to 0.13\%, hashing from
1.17\% to 0.04\%, and policy binding and context/consent/risk evaluation
each fall below 0.05\% at the highest load. Thus, the signed evidence
representation is not the dominant local computational expense. Read
together, Figures~\ref{fig:latency_breakdown} and
\ref{fig:pan_processing_cost} distinguish local PAN computation from the broader authorization delay. Credential validation accounts for most PAN-local processing, whereas evidence encoding, hashing, signing, and Evidence-quorum calculation contribute only small shares. Within the instrumented PAN-side components, these results show that credential validation dominates local processing, whereas evidence encoding, hashing, signing, and local quorum calculation contribute comparatively small shares. The broader authorization cost is therefore associated primarily with distributed evidence collection and PAN-local state processing, rather than with the local construction of signed evidence or the local quorum calculation itself.
\subsection{Policy Activation, Evidence Processing, and Quorum Authorization Results}
\label{subsec:mechanism_results}
Having quantified the cost of operating the distributed authorization path, we now turn to the mechanisms that determine whether that path can progress at all: the state transitions and decision boundaries governing authorization itself. Two mechanisms are central to decentralized sticky policy enforcement: the conditions under which a new sticky policy becomes authorization-active, and the way multiple PANs jointly reach each request-time authorization decision. We examine policy activation first, since PAN evidence is only meaningful with respect to the policy state the participating PANs have actually applied.

\begin{figure}[!t]
\centering
\includegraphics[width=0.95\columnwidth]{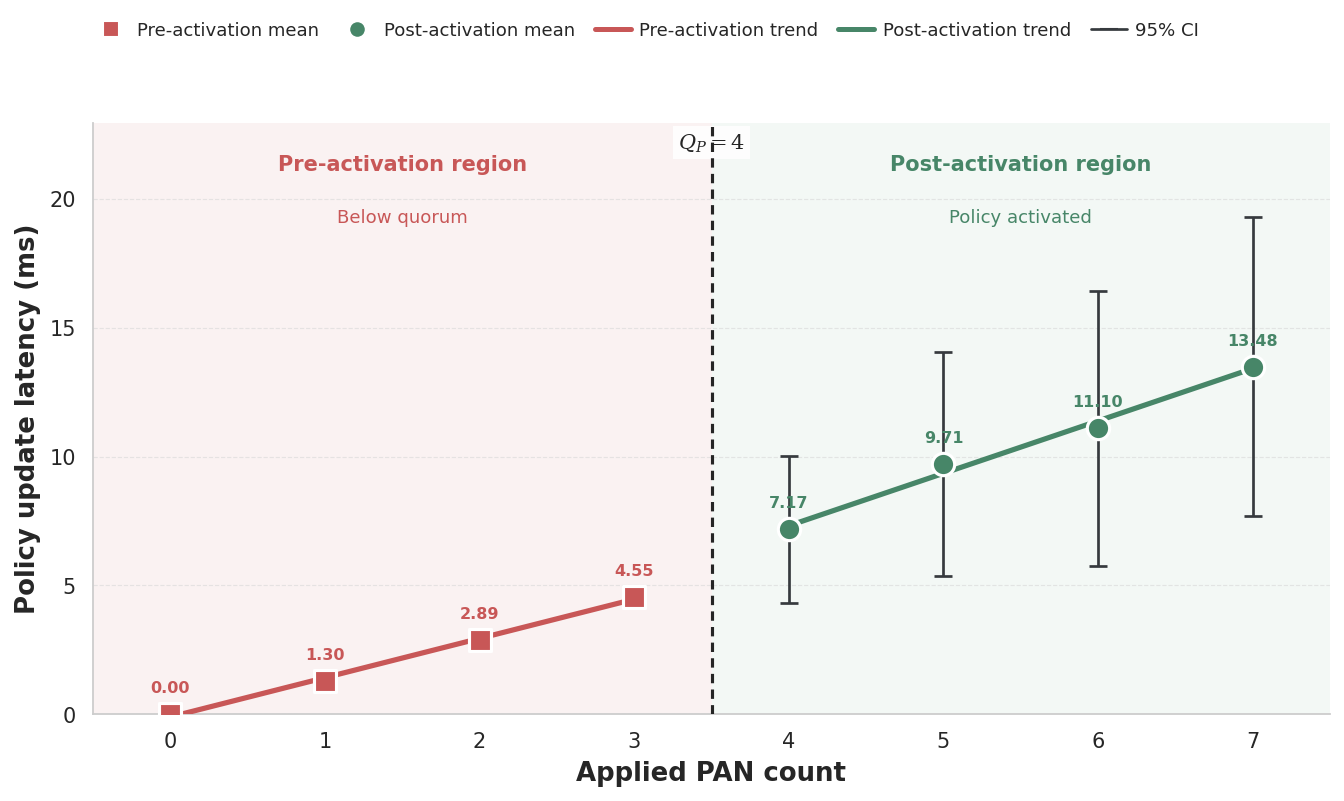}
\caption{Policy-activation boundary for $N=7$ and $Q_P=4$. Error bars show 95\% confidence intervals.}
\label{fig:policy_activation}
\end{figure}

For N=7, policy activation requires Q\_P=4 applied PANs. Figure~\ref{fig:policy_activation} traces this transition as successive PANs apply the committed version. Latency rises steadily but the policy remains inactive through the first three applications, from 0 ms with no applied PAN to 1.30, 2.89, and 4.55 ms as the first, second, and third PANs apply the update. Only when the fourth PAN applies the update, at a measured mean of 7.17 ms, does activation occur. The fifth, sixth, and seventh applications, reaching 9.71, 11.10, and 13.48 ms, continue to raise latency but arrive after the activation boundary has already been crossed and therefore have no further bearing on whether the policy is active.

This transition directly evaluates the activation condition defined in Algorithm \ref{alg:policy_sync}, where a committed policy becomes authorization-active only once at least Q\_P PANs report the same applied version and digest. The measured behavior confirms that condition in practice that the new policy stays inactive until the required PAN majority is reached, which prevents requests from being evaluated against an incompletely applied policy state that could otherwise produce inconsistent authorization decisions across PANs. No single PAN, however early it applies the update, can activate the policy alone, activation is a property of distributed agreement among PANs, not of any individual PAN's action.

Table~\ref{tab:policy_correctness} shows how the policy-update process behaves under different conditions. Valid updates were committed and activated for all three cluster sizes, while repeated and rollback versions were rejected. If Raft did not receive enough acknowledgements, the update was not committed; if too few PANs applied the update, the policy was not activated. Delayed PANs could still apply the policy later. All tested cases produced the expected outcome for $N\in{3,5,7}$, confirming that SPEAR-Q correctly separates policy commitment from policy activation. These cases exercise the validation, Raft-commitment, PAN-application, and activation branches of Algorithm \ref{alg:policy_sync}, confirming that policy commitment and policy activation remain distinct lifecycle states across all evaluated cluster sizes. Together with Figure \ref{fig:policy_activation}, they further show that a delayed PAN can still apply the active policy after the fact, without disrupting or bypassing the majority-based activation requirement. Authorization therefore relies on the majority-approved active sticky policy state, rather than permitting any individual PAN to activate a different policy independently.

\begin{figure}[H]
\centering
\includegraphics[width=0.95\columnwidth]{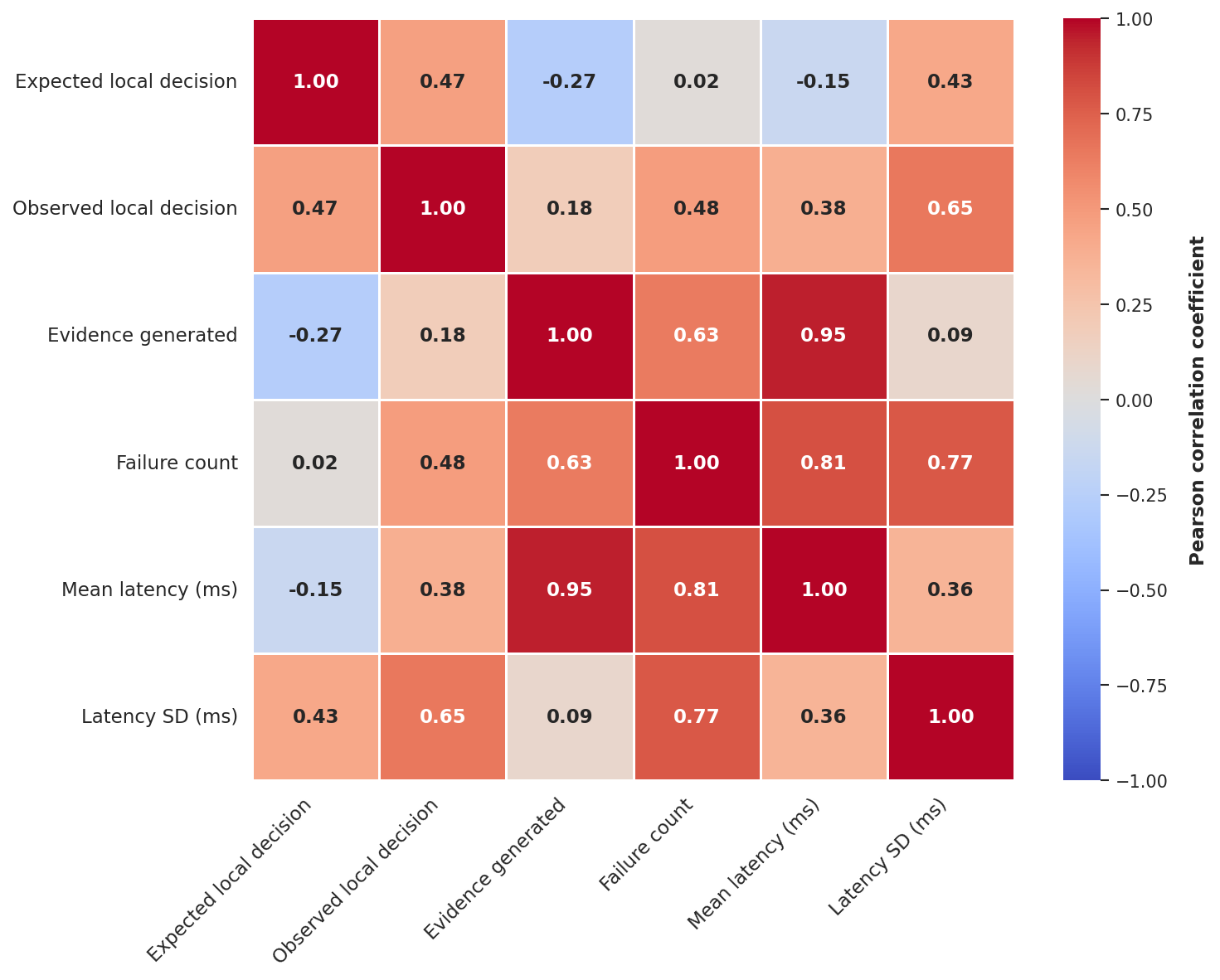}
\caption{Pearson correlation among timing-related evidence-processing measurements, including evidence-generation time, mean request latency, latency variation, and failure count.}
\label{fig:evidence_heatmap}
\end{figure}

Once the sticky policy is active, each request is evaluated against that policy before PAN evidence can contribute to the Evidence quorum, so the deployed authorization path enforces policy at request time for every access request, rather than treating policy attachment as a one-time decision made in advance. This evaluation corresponds to Algorithm \ref{alg:evidence_generation}, where an eligible PAN checks the policy, context, consent, and risk conditions and then constructs, hashes, and signs the resulting authorization evidence. 
Figure~\ref{fig:evidence_heatmap} shows the relationships among the main measurements collected from these evidence-processing cases.
The heatmap shows that time spent on evidence generation is most strongly related to mean latency, with a Pearson correlation coefficient of 0.95, showing a strong association between evidence-generation time and overall request latency. Failure count is also related to mean latency (0.81) and latency variation (0.77), suggesting that failed cases tend to coincide with slower, less predictable processing, while its weaker correlation with evidence generation (0.63) suggests failures are not simply a byproduct of evidence-generation cost alone.

Together, these correlations describe how far requests progress through the evidence-processing path and the cost associated with that progression, but they do not by themselves determine whether a request is authorized. The Evidence-quorum decision is therefore evaluated next, under the compromised-PAN and evidence-loss conditions defined in the threat model in Section~\ref{sec:threat_model}.For $N=7$, authorization requires an Evidence quorum of $Q_E=4$. In the compromised-PAN cases, honest PANs return Deny, while
$m\in\{0,1,2,3\}$ denotes the number of compromised PANs contributing
correctly signed false-Permit evidence. This tests whether fewer than $Q_E$ compromised PANs can cause a Permit. We also reduce the number of reachable PANs below $Q_E$ to test whether authorization remains fail-closed when enough evidence cannot be collected. Table \ref{tab:quorum_authorization} reports both cases.

\begin{table*}[!t]
\caption{Policy-lifecycle validation across PAN cluster sizes.
A check mark indicates that the observed lifecycle outcome matched
the expected behavior.}
\label{tab:policy_correctness}
\centering
\small
\setlength{\tabcolsep}{6pt}
\renewcommand{\arraystretch}{1.16}

\begin{tabular}{|l|c|l|c|c|c|}
\hline
\textbf{Policy condition} &
\textbf{Expected} &
\textbf{Observed state} &
\textbf{$N=3$} &
\textbf{$N=5$} &
\textbf{$N=7$} \\
\hline

Valid update
& Accept
& Committed and activated
& \cmark & \cmark & \cmark \\
\hline

Repeated version
& Reject
& Not activated
& \cmark & \cmark & \cmark \\
\hline

Rollback version
& Reject
& Not activated
& \cmark & \cmark & \cmark \\
\hline

Insufficient Raft acknowledgements
& Reject
& Not committed
& \cmark & \cmark & \cmark \\
\hline

Insufficient PAN quorum
& Reject
& Not activated
& \cmark & \cmark & \cmark \\
\hline

Delayed PAN catch-up
& Accept
& Active; lagging PAN catches up
& \cmark & \cmark & \cmark \\
\hline

\end{tabular}
\end{table*}

 These experiments directly evaluate the evidence-validation and Evidence-quorum decision defined in Algorithm \ref{alg:quorum_authorization}, where only distinct admissible PAN evidence contributes to the decision and a Permit requires at least \(Q_E\) Permit-supporting records.

Table~\ref{tab:quorum_authorization} shows that evidence wait and total latency decrease as \textit{m} increases. This trend is consistent with the SPEAR-Q authorization path, where compromised PANs contributing correctly signed false-Permit evidence do not necessarily complete the same policy, context, consent, and risk-evaluation processing performed by honest PANs before returning an authorization response. As a result, the Resource Verifier may obtain sufficient admissible evidence to determine the authorization outcome earlier as additional compromised responses become available. Importantly, this timing difference affects only how quickly the authorization outcome is reached; authorization safety remains preserved under all evaluated conditions, with no sub-quorum coalition producing a global Permit or releasing the protected resource.

\begin{table*}[!t]
\caption{Quorum authorization results for $N=7$ and $Q_E=4$.
Here, $m$ denotes the number of compromised PANs contributing correctly
signed false-Permit evidence; the second condition varies the number of
reachable PANs. Latency values are reported as mean $\pm$ standard deviation
across independent repetitions.}
\label{tab:quorum_authorization}
\centering
\scriptsize
\setlength{\tabcolsep}{5pt}
\renewcommand{\arraystretch}{1.14}

\begin{tabular}{|l|l|c|c|c|c|c|c|c|}
\hline
\textbf{Condition} &
\textbf{Setting} &
\textbf{Reps.} &
\textbf{\shortstack{Evidence wait\\(ms)}} &
\textbf{\shortstack{Mean latency\\$\pm$ SD (ms)}} &
\textbf{95\% CI (ms)} &
\textbf{Expected} &
\textbf{Observed} &
\textbf{Release} \\
\hline

\multirow{4}{*}{False-Permit support}
& $m=0$ & 5
& 14.51
& $25.28\pm0.86$
& $[24.22,\,26.35]$
& Deny & Deny & No \\
\cline{2-9}

& $m=1$ & 16
& 12.30
& $23.09\pm2.15$
& $[21.94,\,24.23]$
& Deny & Deny & No \\
\cline{2-9}

& $m=2$ & 8
& 8.86
& $17.63\pm0.95$
& $[16.83,\,18.42]$
& Deny & Deny & No \\
\cline{2-9}

& $m=3$ & 30
& 7.50
& $14.99\pm2.52$
& $[14.05,\,15.93]$
& Deny & Deny & No \\
\hline

\multirow{4}{*}{Reachable PANs}
& 3 & 6
& 3.75
& $9.49\pm0.42$
& $[9.04,\,9.93]$
& Deny & Deny & No \\
\cline{2-9}

& 2 & 28
& 2.71
& $9.17\pm1.18$
& $[8.71,\,9.63]$
& Deny & Deny & No \\
\cline{2-9}

& 1 & 29
& 1.63
& $7.90\pm1.03$
& $[7.51,\,8.29]$
& Deny & Deny & No \\
\cline{2-9}

& 0 & 8
& 0.00
& $5.62\pm0.32$
& $[5.35,\,5.89]$
& Deny & Deny & No \\
\hline

\end{tabular}
\end{table*}

\begin{figure}[!t]
\centering
\includegraphics[width=1.0\columnwidth]{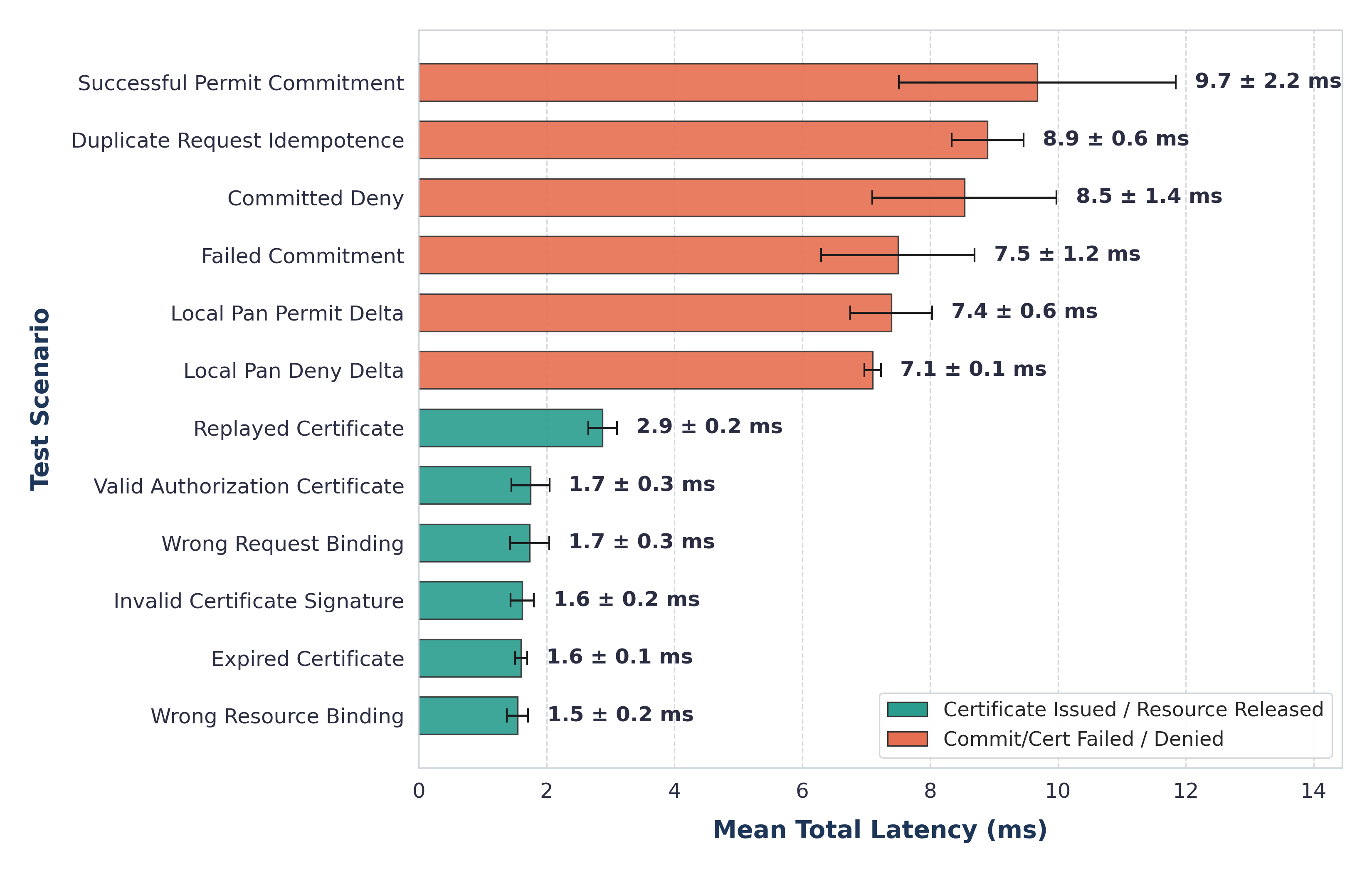}
\caption{Post-authorization processing latency}
\label{fig:post_authorization}
\end{figure}

After confirming the Evidence-quorum behaviour, the evaluation next examines the cost of post-authorization processing. This stage corresponds to the post-authorization portion of Algorithm \ref{alg:quorum_authorization}, covering authorization-record commitment, PAN-local risk-state update, certificate issuance, and conditional resource release. Figure~\ref{fig:post_authorization} shows that certificate-related operations remain lightweight, with most cases between 1.5 and 1.7~ms and replay handling reaching 2.9~ms. Commitment- and state-related operations are higher, ranging from 7.1~ms for a local-PAN Deny update to 9.7~ms for the successful Permit-commitment path, while duplicate-request handling reaches 8.9~ms and committed Deny 8.5~ms. This increase reflects the additional state-management work required after authorization. However, these costs remain small compared with the PAN evidence wait and PAN-local risk-update costs observed under high concurrency in figure~\ref{fig:latency_breakdown}, showing that post-authorization certificate processing is not the main source of high-load latency.

\subsubsection{Comparative Evaluation}

The closest prior approaches decentralize different parts of the sticky policy and authorization path. Rizzardi et al.~\cite{rizzardi2022securing} decentralize sticky policy management, Zichichi et al.~\cite{zichichi2020personal} require threshold participation for protected-data release, and Alshahrani et al.~\cite{alshahrani2026decentralized} come closest to decentralized request-time authorization through independent validator decisions. Table~\ref{tab:closest_comparison} summarizes the properties most directly related to the mechanisms evaluated in SPEAR-Q.

\begin{table}[!t]
\caption{Comparison with the closest sticky policy and decentralized authorization approaches. SP: Sticky Policy support; DA: Decentralized Request-Time Authorization; QT: Quorum or Threshold authorization; PA: Policy Activation mechanism; CB: explicit Compromise Bound.}
\label{tab:closest_comparison}
\centering
\small
\renewcommand{\arraystretch}{1.1}

\resizebox{\columnwidth}{!}{
\begin{tabular}{|l|c|c|c|c|c|}
\hline
\textbf{Approach} &
\textbf{SP} &
\textbf{DA} &
\textbf{QT} &
\textbf{PA} &
\textbf{CB} \\
\hline

Rizzardi et al.~\cite{rizzardi2022securing}
& $\checkmark$ & $\times$ & $\times$ & $\times$ & $\times$ \\
\hline

Alshahrani et al.~\cite{alshahrani2026decentralized}
& $\times$ & $\checkmark$ & $\checkmark$ & $\times$ & $\checkmark$ \\
\hline

Zichichi et al.~\cite{zichichi2020personal}
& $\times$ & $\times$ & $\checkmark$ & $\times$ & $\times$ \\
\hline

SPEAR-Q
& $\checkmark$ & $\checkmark$ & $\checkmark$ & $\checkmark$ & $\checkmark$ \\
\hline

\end{tabular}}
\end{table}

Rizzardi et al. decentralize the sticky policy lifecycle rather than the request-time authorization decision, while Zichichi et al. distribute the authority required for threshold-controlled data release. Alshahrani et al. provide the closest request-time analogue because multiple validators contribute to the same authorization decision. In contrast, SPEAR-Q combines persistent sticky policy enforcement, decentralized request-time authorization, majority-based policy activation, Evidence-quorum authorization, and an explicit compromised-node bound within a single authorization path. Its evaluation examines these mechanisms under increasing cluster size and concurrency, sub-quorum false-Permit evidence, and insufficient evidence.

The comparison is deliberately structural rather than numerical, since differences in architecture, workload, deployment, and measurement scope prevent a defensible comparison of absolute latency or throughput. The cited approaches therefore provide context for the SPEAR-Q results rather than a matched performance baseline; isolating the cost attributable specifically to decentralization would require an otherwise identical centralized PDP implementation under the same experimental conditions. The implications of the measured SPEAR-Q results are discussed next.
\subsection{Discussion}
\label{subsec:discussion}

This discussion integrates the performance and mechanism results to identify the main sources of authorization cost, assess whether the defined security properties are preserved, and clarify the scope of the findings.

\subsubsection{Sources of Authorization Latency Under Load}
The performance results reveal a clear trade-off between decentralized authorization and scalability as cluster size and concurrent load increase. All 9,000 performance requests completed without failure or timeout, showing that the observed high-load degradation reflects saturation rather than request loss or execution failure; for \(N=3\), throughput at \(C=100\) was approximately 24\% below its measured peak, while \(N=5\) and \(N=7\) reached their peak throughput earlier at \(C=10\). 

The latency decomposition further identifies where this scaling pressure accumulates. For \(N=7\), PAN evidence waiting increases from 26.0\% of end-to-end latency at \(C=1\) to 44.0\% at \(C=100\), while PAN-local risk processing remains substantial at 35.9\% and 34.8\%, respectively; in contrast, the relative contribution of authorization-record commitment decreases from 17.1\% to 2.9\%. The fixed-\(N=3\) analysis shows the same change under increasing concurrency: at \(C=100\), PAN evidence waiting and PAN-local risk update account for 36.3\% and 33.7\% of end-to-end latency, whereas commitment contributes only 4.5\%. Gateway validation remains at or below 1\%, and sticky policy attachment and Evidence-quorum calculation remain negligible relative to the distributed processing stages. These percentages should not be interpreted as a complete decomposition because Figure~\ref{fig:latency_breakdown} retains a residual component representing latency outside the explicitly instrumented boundaries.

On the PAN side, credential validation becomes the dominant measured local operation, increasing from 90.48\% of PAN processing at \(C=1\) to 99.78\% at \(C=100\), while signing falls from 5.37\% to 0.13\%, hashing from 1.17\% to 0.04\%, policy binding from 1.37\% to 0.03\%, and context/consent/risk evaluation from 1.37\% to 0.02\%. 

Together, these results point to distributed evidence collection and credential validation as the main measured scaling pressures, rather than evidence construction or local quorum calculation. This distinction matters because it places the dominant cost with authorization-input validation, not with the sticky policy or quorum mechanisms that SPEAR-Q introduces. Evidence construction, hashing, signing, and quorum evaluation all stay comparatively small across the entire evaluated range.

\subsubsection{Mechanism Robustness and Security Properties}
\textbf{Policy Activation}
The mechanism experiments first examine whether SPEAR-Q preserves the policy-activation boundary under delayed or incomplete policy application. For \(N=7\), policy activation requires \(Q_P=4\), corresponding to 57.1\% of the PAN cluster, and the policy remains inactive while only three PANs (42.9\%) have applied the committed state; Table~\ref{tab:policy_correctness} further shows that valid updates, rollback and repeated versions, insufficient Raft acknowledgements, insufficient PAN application, and delayed catch-up all produce the expected lifecycle outcome across \(N\in\{3,5,7\}\). These results show that policy commitment and policy activation remain distinct lifecycle stages and that no individual PAN can independently activate a new policy state.

\textbf{Evidence-Quorum Security}
The Evidence-quorum experiments then evaluate whether the authorization threshold remains intact under PAN compromise and evidence loss. For \(N=7\), authorization requires \(Q_E=4\), while the largest tested false-Permit coalition contains only three compromised PANs (42.9\% of the cluster). Across all evaluated compromised-PAN and insufficient-evidence conditions, the observed outcome remained Deny with no resource release, supporting the implemented \(m<Q_E\) compromise bound and fail-closed behaviour without reducing the quorum threshold. Thus, neither a sub-quorum coalition nor reduced PAN availability lowers the authorization threshold under the evaluated conditions. 

\textbf{Timing Behaviour and Measurement Precision}
The timing measurements provide a separate view of how these adversarial conditions affect request processing without changing the authorization boundary. Evidence-processing measurements also show a strong association between evidence generation and mean latency (\(r=0.95\)), while failure count correlates with mean latency (\(r=0.81\)) and latency variation (\(r=0.77\)); these relationships describe processing behaviour rather than authorization correctness. Evidence wait and total latency decrease as the number of compromised PANs \(m\) increases. This pattern is consistent with compromised PANs returning correctly signed false-Permit evidence without incurring the same policy, context, consent, and risk-evaluation processing performed by honest PANs. As more such responses become available, the Resource Verifier can collect sufficient admissible evidence and resolve the request sooner. This timing difference affects how quickly the decision is reached rather than changing the authorization outcome. The reported relative 95\% confidence-interval half-widths range from approximately 4.2-6.3\% for the false-Permit cases and 4.7-5.0\% for the reachable-PAN cases. The 5\% precision target is therefore not achieved uniformly across all timing measurements. This uncertainty affects the precision of the measured latency rather than the categorical authorization outcome, which remained Deny with no resource release in every tested case.

\subsubsection{Limitations and Scope}

The results reported above should be interpreted within the specific boundaries of this evaluation. The 40,000-request workload comprises 50\% legitimate requests, 35\% authorization violations, and 15\% protocol and policy attacks; MIMIC-IV supplies the protected healthcare objects, while authorization parameters are assigned by the experiment scripts. The deployment itself is limited to physically separated hosts on a single isolated subnet, with cluster size capped at \(N\leq7\) and concurrency capped at \(C\leq100\).

The Policy Issuer, Resource Verifier, and Provider are trusted, Raft is considered under crash and communication faults rather than Byzantine behaviour, the compromised-PAN guarantee applies only for \(m<Q_E\), and the risk-update values are prototype parameters rather than a calibrated behavioural-risk model. Furthermore, no matched centralized PDP baseline is evaluated, so the evaluation can characterize SPEAR-Q's absolute performance and internal cost distribution, but cannot quantify the additional latency or throughput cost attributable specifically to decentralization relative to centralized authorization. 

Within these limits, the combined evidence supports the central claim that sticky policy activation and request-time authorization can be distributed across multiple PANs while preserving the defined majority-based activation, fail-closed Evidence-quorum, and conditional-release properties under the evaluated conditions.
\section{Conclusion}
\label{sec:conclusion}

This paper addressed the dependence of sticky policy systems on centralized request-time authorization by introducing SPEAR-Q, a decentralized framework in which independent Policy Authority Nodes (PANs) evaluates policy, context, consent, and risk conditions and produce signed authorization evidence. A strict-majority Evidence quorum determines the global authorization decision, while Raft is used for policy-update and authorization-record commitment rather than request-time decision making. In this way, SPEAR-Q distributes authorization authority across multiple PANs while retaining majority-based policy activation and conditional resource release.

SPEAR-Q was deployed across physically separated hosts on the Airbus Cyber Security Simulation Platform and evaluated using MIMIC-IV-derived healthcare workloads. Latency increased and throughput saturated as cluster size and concurrency grew, while all 9,000 performance requests completed without failure or timeout. PAN evidence waiting and PAN-local processing dominated under load; within PANs, zero-trust credential validation was the main cost, while Evidence-quorum calculation and Raft commitment added comparatively little overhead. Mechanism experiments showed that a committed policy became the active policy state only after the required PAN majority applied it, while sub-quorum false-Permit evidence could not produce a global Permit and insufficient evidence resulted in fail-closed authorization with no resource release. These results support SPEAR-Q’s majority-based activation, Evidence-quorum authorization, and conditional-release properties under the evaluated conditions.

The conclusions remain bounded by the experimental setting. The deployment was limited to an isolated network with \(N\leq7\) and \(C\leq100\), the workload was healthcare-derived with authorization parameters generated by the experiment framework, trusted components were assumed, the compromised-PAN guarantee applies only for \(m<Q_E\), and no matched centralized PDP baseline was evaluated. Future work will examine broader workloads and deployment environments, larger and geographically distributed PAN configurations, matched comparisons with centralized and representative decentralized authorization approaches under equivalent experimental conditions, and finer-grained profiling of the credential-validation path identified as the principal PAN-side scaling bottleneck.

\section*{Acknowledgments}
This work was supported by Research Ireland and the CONNECT Research Centre under Grant No. 13/RC/2077\_P2.


\bibliographystyle{IEEEtran}

\end{document}